\documentclass[twocolumns,structabstract]{aa}
\usepackage{amsmath}
\usepackage{graphicx}
\usepackage{txfonts}
\usepackage{natbib}
\usepackage{mathtools}
\usepackage{color}
\usepackage{hyperref}

\bibpunct{(}{)}{;}{a}{}{,} 

\hypersetup{draft}
\begin{document}

\title{Transit Ly-$\alpha$ signatures of terrestrial planets \\in the habitable zones of M dwarfs}

   \author{K.~G.~Kislyakova
          \inst{1,2}
          \and
          M.~Holmstr\"{o}m
          \inst{3}
          \and
          P.~Odert
          \inst{2}  
          \and          
          H.~Lammer
          \inst{2}                              
          \and                    
          N.~V.~Erkaev
          \inst{4,5}                               
          \and
          M.~L.~Khodachenko
          \inst{2,6}                               
          \and
          I.~F.~Shaikhislamov
          \inst{7}                               
          \and
          E.~Dorfi
          \inst{1}
          \and
          M.~G\"{u}del
          \inst{1}}


   \institute{University of Vienna, Department of Astrophysics,  T\"{u}rkenschanzstrasse 17, A-1180 Vienna, Austria
		 \and   
            Space Research Institute, Austrian Academy of Sciences, Schmiedlstrasse 6, A-8042 Graz, Austria
         \and
            Swedish Institute of Space Physics, PO Box 812, SE-98128 Kiruna, Sweden
         \and
            Institute of Computational Modelling, Siberian Division of Russian Academy of Sciences, 660036 Krasnoyarsk, Russian Federation
         \and
            Siberian Federal University, Krasnoyarsk, Russian Federation
         \and
            Skobeltsyn Institute of Nuclear Physics, Moscow State University, Moscow, Russian Federation
         \and
         Institute of Laser Physics SB RAS, Novosibirsk, Russian Federation}

   \date{Received \today}

\abstract{}
{We modeled the transit signatures in the Lyman-alpha (Ly$\alpha$) line of a putative Earth-sized planet orbiting in the habitable zone (HZ) of the M~dwarf GJ~436. We estimated the transit depth in the Ly$\alpha$ line for an exo-Earth with three types of atmospheres: a hydrogen-dominated atmosphere, a nitrogen-dominated atmosphere, and a nitrogen-dominated atmosphere with an amount of hydrogen equal to that of the Earth. For all types of atmospheres, we calculated the in-transit absorption they would produce in the stellar Ly$\alpha$ line. We applied it to the out-of-transit Ly$\alpha$ observations of GJ~436 obtained by the Hubble Space Telescope and compared the calculated in-transit absorption with observational uncertainties to determine if it would be detectable. To validate the model, we also used our method to simulate the deep absorption signature observed during the transit of GJ~436b and showed that our model is capable of reproducing the observations.  }
{We used a direct simulation Monte Carlo (DSMC) code to model the planetary exospheres. The code includes several species and traces neutral particles and ions. It includes several ionization mechanisms, such as charge exchange with the stellar wind, photo- and electron impact ionization, and allows to trace particles collisions. At the lower boundary of the DSMC model we assumed an atmosphere density, temperature, and velocity obtained with a hydrodynamic model for the lower atmosphere. }
{We showed that for a small rocky Earth-like planet orbiting in the HZ of GJ~436 only the hydrogen-dominated atmosphere is marginally detectable with the Space Telescope Imaging Spectrograph (STIS) on board the Hubble Space Telescope (HST). Neither a pure nitrogen atmosphere nor a nitrogen-dominated atmosphere with an Earth-like hydrogen concentration in the upper atmosphere are detectable. We also showed that the Ly$\alpha$ observations of GJ~436b can be reproduced reasonably well assuming a hydrogen-dominated atmosphere, both in the blue and red wings of the Ly$\alpha$ line, which indicates that warm Neptune-like planets are a suitable target for Ly$\alpha$ observations. Terrestrial planets, on the other hand, can be observed in the Ly$\alpha$ line if they orbit very nearby stars, or if several observational visits are available.  }{}


\titlerunning{Transit Ly-$\alpha$ signatures of terrestrial planets}
\authorrunning{K.G.~Kislyakova et al.}
\maketitle

\section{Introduction}
\label{intro}

Low-mass M~dwarfs are the most numerous stars in the Universe and are often hosts to small rocky planets of a size and mass similar to the Earth. These planets are interesting targets for modern planetology, because conditions in the systems of M~dwarfs differ significantly from those in the planetary systems around G~stars, thus raising questions about the evolution of such planets (e.g., \citealp{Scalo07}).

Lyman-alpha (Ly$\alpha$) transit observations are a powerful tool of characterizing the planetary atmospheres. For hot Jupiters, in a number of cases these observations can be used as an indicator of escape rates from hydrogen-dominated atmospheres (e.g., \citealp{VM03,Lecavelier12,BL13,Khodachenko17}), atmosphere properties \citep{BJ07} and planetary magnetic moments and stellar winds \citep{K14b,Vidotto17}. Besides hot Jupiters, a variety of smaller exoplanets has also been observed in Ly$\alpha$, including planets for which a deep absorption signature indicating a presence of a dense and extensive hydrogen envelope has been detected, such as GJ~436 \citep{Kulow14,Ehrenreich15,Lavie17} and GJ~4370b \citep{B18b}, and planets for which no excess absorption in comparison to the visible light has been found \citep{Ehrenreich12,B17b}. Different character of observed in-transit signals of exoplanets indicates that Ly$\alpha$ observations can be used as a powerful tool not only to detect a presence of a hydrogen envelope around an exoplanet, but also its shape and density. Ly$\alpha$ observations can also give us hints on other parameters such as planetary magnetic moment and stellar wind density and velocity, and the type of interaction between planetary atmosphere and stellar wind \citep{Khodachenko15,Khodachenko17,Shaikhislamov16}. However, one should keep in mind that interpreting the Hubble Space Telescope (HST) Ly$\alpha$ observations can be very challenging (for instance, see discussion in \citealp{BJ07} and \citealp{VM08}, and in other papers cited above).

At present, Ly$\alpha$ observations can be obtained only by the Hubble Space Telescope, as this is the only space observatory equipped with an UV instrument. In the future, other space observatories such as LUVOIR \citep{France17} and WSO UV \citep{Boyarchuk16} may provide us with more precise observations. 

In this paper, we estimated the transit depth in the Ly$\alpha$ line of a terrestrial planet orbiting in a habitable zone (HZ) of an M~star. Our ultimate goal was to examine if planets with nitrogen-dominated atmospheres showed an excess absorption in the Ly$\alpha$ line due to hydrogen in their exospheres originating from the charge exchange between incoming stellar wind protons and exospheric neutrals. If these signatures would be detectable, the observations in the Ly$\alpha$ line could be used as a characterization tool not only for hydrogen-dominated planets, but also for exoplanets with secondary atmospheres. We have used Ly$\alpha$ observations of an M~dwarf GJ~436 of a spectral type M2.5, which hosts a transiting hot Neptune GJ~436b \citep{Ehrenreich15,Lavie17}. This planet has been observed to produce an immense in-transit Ly$\alpha$ absorption, thus allowing for a sophisticated modeling of its hydrogen envelope and other parameters of the system \citep{B15,B16,Vidotto17,Loyd17}. For this study, we have selected out-of-transit observations of GJ~436 published by \citet{Lavie17}, which we used as stellar background spectrum for our modeling. We modeled atomic coronae surrounding an Earth-mass planet orbiting in the habitable zone (HZ) of GJ~436 with a nitrogen and a hydrogen-dominated atmosphere and then calculated absorption in the Ly$\alpha$ line which would be produced by a transit of such an exospheric cloud. Then, we imposed this calculated absorption on the observed Ly$\alpha$ profile of GJ~436 and compared it with the observational uncertainties. This method allowed us to conclude if planets with different types of atmospheres could be observable in the Ly$\alpha$ line by the HST. We performed the modeling for several XUV (X-ray and extreme ultraviolet) enhancement factors: three times the modern solar flux at Earth (which is the nominal XUV flux in the middle of the HZ of GJ~436 according to observations by \citealp{Ehrenreich11}), seven, ten, and twenty times. For the case of 3 XUV, we include an additional case of a nitrogen-dominated atmosphere with a hydrogen content equal to the one at the Earth's exobase.

This paper is organized as follows: in Section~\ref{sec_model}, we describe our models and available codes we use for the simulations. To validate our method, in Section~\ref{sec_GJ436b} we apply our model to GJ~436b and show that our model can reproduce a deep observed transit signature of this planet. In Section~\ref{sec_parameters}, we describe simulation parameters and inputs. Section~\ref{sec_results} summarizes our results. Section~\ref{sec_disc} describes possible applications of the results to observation planning and compares our results with the literature, and, finally, Section~\ref{sec_conc} summarizes our conclusions.

\section{Model description}
\label{sec_model}

\subsection{The star. GJ~436}

GJ~436 is an M~dwarf of spectral class M2.5 with a mass of 0.452~M$_\odot$ and a radius of 0.464~R$_\odot$. We have selected GJ~436 as a target because for this star Ly$\alpha$ in-transit and out-of-transit observations with the Space Telescope Imaging Spectrograph (STIS) on board the Hubble Space Telescope (HST) are available. Due to its relatively close distance to the Earth of only 10.2~pc, this observation is of a very high quality and has a high signal-to-noise ratio. Transit observations of GJ~436b by the HST revealed an immense absorption in the stellar Ly$\alpha$ line, especially in its blue wing \citep{Ehrenreich15}. Later visits with longer exposure times confirmed the presence of a deep in-transit, egress, and ingress absorption, with the mid-transit signature being the deepest \citep{Lavie17}.   This indicates a presence of a massive hydrogen cloud around this planet, which obscures almost half of the emission in the stellar Ly$\alpha$ line (the absorption during a transit is $\approx$56\%). Numerical modeling confirms that GJ~436b is surrounded by a big cloud of atomic hydrogen with a complex structure \citep{B15,B16,Loyd17,Shaikhislamov18}.

In this article, we aimed at investigating if a smaller planet located further from GJ~436 than GJ~436b, namely, an Earth-like planet in the habitable zone of this star with the orbital distance of 0.24~au, could also produce a detectable Ly$\alpha$ signature. For this purpose, we used out-of-transit observations of GJ~436b and calculated the Ly$\alpha$ absorption caused by a transit of such a planet in the HZ. M~dwarfs such as GJ~436 are very numerous in the solar neighborhood, and therefore this study can be used for predicting the amount of Ly$\alpha$ absorption small rocky planets orbiting them can produce.


\subsection{Atmosphere parameters}
\label{sec_atmpar}

Even if a planet is similar to Earth in terms of mass and size, it does not necessarily have an atmosphere similar to our planet, in other words, dominated by nitrogen. The amount of hydrogen the planet accumulates depends on its formation history. If a planet grows to a mass close to Earth's mass before the protoplanetary nebula dissipates, it can accumulate a dense hydrogen envelope \citep{Stoekl16}. After the nebula dissipates, this envelope will partly escape due to enhanced thermal escape \citep{Owen16a,Fossati17}, however, Earth-like planets in the HZs can still keep a significant fraction of it. If the hydrogen envelope does not disappear during the first tens of Myr after the planet has been released from the nebula, it can still escape afterwards, both by thermal (e.g., \citealp{Erkaev13,L14,Luger15,Johnstone15b,Kubyshkina18}) and non-thermal (e.g., \citealp{K13,Airapetian17}) escape mechanisms. Of course, the exact evolutionary path of a planet strongly depends on the stellar activity evolution and on the initial amount of hydrogen the planet accumulates. 

On the other hand, if a planet never accumulates enough hydrogen, or if the star is active enough early on, it is possible that a secondary atmosphere forms. In this article, we focus on nitrogen-dominated atmospheres. One should keep in mind that these atmospheres may not be stable around active M~dwarfs, because a nitrogen-dominated atmosphere readily expands if exposed to strong XUV radiation \citep{Tian08}, which leads to strong non-thermal losses and, consequently, to a rapid loss of the whole atmosphere \citep{Lichtenegger10,Airapetian17}. Due to their slow evolution, M~stars can keep high levels of XUV radiation in their HZs for a longer time in comparison to solar-like stars \citep{West08}, which can further enhance non-thermal losses. The atmospheres of planets orbiting active low mass M~dwarfs may be replenished by extreme volcanic activity due to tidal or induction heating \citep{Driscoll15,K17,K18}.  A volcanically active planet would eject gases composed mainly of SO$_2$, CO$_2$, H$_2$O, and S$_2$. The exact composition depends on the redox state and volatile abundances in planetary interiors, and also on the surface pressure \citep{Gaillard14}, but a volcanically produced nitrogen-dominated atmosphere seems unlikely. At Earth, both life and plate tectonics support the nitrogen-dominated atmosphere \citep{L18}. 

It seems reasonable to assume that Earth-like planets in the HZs of M~stars can have both types, hydrogen-dominated and secondary, atmospheres. Observations confirm the presence of planets with and without a hydrogen envelope, which can be concluded about based on their average density (e.g., \citealp{Lopez12,Ginzburg16,Fossati17}).

\noindent\textbf{Hydrogen atmospheres.} To obtain the parameters at the exobase of a hydrogen-dominated atmosphere, we applied a hydrodynamic code, which is a 1D upper atmosphere radiation absorption and hydrodynamic escape model that takes into account ionization, dissociation and recombination \citep{Erkaev16,Erkaev17}. The code provides mass-loss rates and atmospheric structure under different radiation conditions, and has been successfully applied multiple times to exoplanets (e.g., \citealp{Erkaev13,Erkaev15,L14,L16}). 

The code calculates the absorption of the stellar XUV flux by the thermosphere and solves the hydrodynamic equations for mass, momentum and energy conservation, and the continuity equations for neutrals and ions (both atoms and molecules). The model also accounts for dissociation, ionization, recombination and Ly$\alpha$ cooling. The quasi-neutrality condition determines the electron density. The model does not self-consistently calculate the ratio of the net local heating rate to the rate of the stellar radiative energy absorption. For this reason, we assumed a heating efficiency of 15 per cent, which is a mean value for a hydrogen-dominated atmosphere according to kinetic studies solving the Boltzmann equation \citep{Shematovich14}. We run the model for 3, 7, 10, and 20 XUV and took the values for the number density, temperature, and gas outflow velocity at the exobase as an inner boundary input for the DSMC model.

\noindent\textbf{Nitrogen atmospheres.} For nitrogen-dominated atmospheres, we adopted the parameters from the literature, namely, from the paper by \citet{L08} based on the results by \citet{Tian08}, who studied the response of Earth's atmosphere to enhanced levels of XUV radiation using a hydrodynamic atmosphere model. \citet{Tian08} and \citet{L08} have shown that at high XUV fluxes, an Earth-like nitrogen-dominated atmosphere with a present atmospheric level of CO$_2$ (1~PAL) expands to high altitudes and is dominated by atomic nitrogen. Venus, on the other hand, despite being exposed to a higher XUV flux than Earth, has a much less expanded atmosphere and a cooler exobase temperature due to the atmospheric composition dominated by CO$_2$ (96\% CO$_2$; e.g., \citealp{L08}). High sensitivity of nitrogen-dominated atmospheres to stellar XUV flux is confirmed by other authors, who applied a hydrodynamic code combined with a sophisticated atmospheric chemistry model \citep{Johnstone18}. Similar to the hydrogen-dominated atmospheres, we adopted the atmospheric parameters for 3, 7, 10, and 20 XUV \citep{L08}. While such radiation levels are typical mostly for young G dwarfs, M dwarfs often exhibit higher short wavelength radiation levels in their HZs even at older ages (e.g., \citealp{West08,France16,Youngblood16}). At present, GJ~436 is a relatively quiet M~dwarf with an XUV flux in the HZ of only 3~XUV (see Section~\ref{subsec_Petra}), however, this star could have produced a higher level of XUV radiation in the past. 

We note that the nitrogen atmosphere can possibly escape within geologically short times even at 10 and 20~XUV \citep{Lichtenegger10}. This escape may be even stronger for non-magnetized planets such as those we consider in this article, especially taking into account that the stellar wind in the HZs of M~dwarfs may be much denser than the solar wind at 1~AU, due to the proximity of the HZs of M~dwarfs to their hosts. However, we still included these cases in our article to check if short-living nitrogen atmospheres may be detectable in the Ly$\alpha$ line, even though nitrogen-dominated atmospheres may not be stable for a geologically long time. An increased amount of CO$_2$, which is a sufficient coolant in the thermosphere, may protect the atmosphere from expansion and enhanced non-thermal escape \citep{Lichtenegger10,Johnstone18}, but modeling of different compositions of atmospheres is beyond the scope of the present article.

We adopted the inner boundary for our DSMC simulations at the exobase, at a radius $R_{\rm ib}$, with a temperature, $T_{\rm exo}$. The values of $R_{\rm ib}$ and $T_{\rm exo}$ were adopted from \citet{L08}. Since the densities were directly not available in this article, we estimated them from the temperature profiles as 

\begin{equation}
n_{\rm exo} = n_{\rm ib} = \frac{1}{\sigma H},
\end{equation}
where $H$ is the local scale height and $\sigma$ is the collision cross section. The local scale height is determined as $H = k_B T_{\rm exo} / m g_{\rm exo}$, where $k_B$ is the Boltzmann constant, $m$ is the mass of an atom of the main species in the upper atmosphere (in this case, nitrogen), and $g_{\rm exo}$ is the local gravitational acceleration at the exobase. We assumed that nitrogen atoms were the only component of the upper atmosphere, which yielded a slight overestimate of their number density as \citet{Tian08} have also taken into account other atmospheric constituents such as oxygen and others. However, as we showed below, even for a slightly overestimated density a nitrogen atmosphere does not produce any detectable Ly$\alpha$ signature, therefore, our results are not influenced by this factor.

To estimate collision cross sections between neutral species, we used a simplified approach following \citet{Atkins00}
\begin{equation}
	\sigma_{a-b} = \pi (R_a + R_b)^2,
\end{equation}
where $\sigma_{a-b}$ is the elastic collision cross section between the two species, and $R_a$ and $R_b$ are the atomic radii of corresponding species. For the species included in the simulations, we obtained (in m$^2$)
\begin{equation}
\begin{split}
	\sigma_{N-N} = 3.9 \times 10^{-20}, \\
	\sigma_{H-H} = 3.5 \times 10^{-20}, \\
	\sigma_{H-N} = 3.7 \times 10^{-20}.
\end{split}
\end{equation}
We did not take into account dependence of collisional cross section on particles velocity. Elastic collisions with and between ions were not included in our model, however, we included charge exchange collisions between stellar wind protons and atmospheric neutrals.


\noindent\textbf{Nitrogen-dominated atmosphere with an Earth-like hydrogen amount.} As an additional test, we calculated also Ly$\alpha$ absorption of a nitrogen-dominated atmosphere at 3~XUV assuming the presence of an additional hydrogen content in the planetary exosphere, equal to the modern terrestrial level with the density at the exobase of $7.0\times10^{10}$~m$^{-3}$ \citep{K13}. This hydrogen density is in agreement with the observations of the Earth exosphere by the IBEX satellite \citep{Fuselier10}.

\subsection{Stellar wind}
\label{sec_sw}

Stellar wind interacts in various ways with planetary atmospheres. It compresses the magnetospheres, in case of planets with intrinsic magnetic fields, and compresses and penetrates planetary ionospheres of non-magnetized planets. Stellar wind is a powerful ionization source due to electron impact ionization and charge exchange between stellar wind particles and atmospheric neutrals. As a result of a charge exchange reaction, an Energetic Neutral Atom (ENA) and a slow ion of planetary origin are produced, because the electron is transferred from the planetary neutral particle to the stellar wind proton, which keeps its energy and velocity. Simulations show that stellar wind can penetrate deep into the atmospheres of non-magnetized planets \citep{Cohen15}, and lead to intense atmospheric erosion and shape the hydrogen cloud surrounding the planet \citep{K14,Vidotto17,Shaikhislamov16}. A similar interaction pattern is also observed at Mars and Venus \citep{Galli08,Galli08b,Lundin11}. If charge exchange occurs between heavy ions of the stellar wind and neutral atoms from an atmosphere, a powerful secondary X-ray emission is generated \citep{K15}. However, in this article we disregarded the effects of heavy ions and focused on charge exchange between stellar wind protons and atmospheric neutrals, which generates ENAs. These atoms keep the velocity of a high energetic stellar wind proton and contribute to the absorption in the wings of the Ly$\alpha$ line. Due to the fact that the electron is transferred to ENAs from an atmospheric neutral atom, this process also contributes to overall ionization in the planetary exosphere. Since charge exchange takes place outside the planetary magnetosphere or ionosphere boundary, these newly produced ions can be easily picked up by the stellar wind and thus also contribute to the non-thermal losses. ENAs which impact the planetary atmosphere can also deposit their energy there and contribute to the atmosphere's energy budget, but their influence on thermal escape has been estimated to be negligible \citep{Lichtenegger16}.

In this article, we adopted the stellar wind data from \cite{K13}. \cite{K13} applied a Versatile Advection Code \citep{Toth96} to model the plasma environment of a GJ436-like star. The model includes a self-consistent Parker-type corotating magnetic field and is based on the solution of the set of the ideal nonresistive nonrelativistic magnetohydrodynamic equations, and yields a self-consistent expanding stellar wind plasma flow. In the middle of the HZ (at 0.24~au), stellar wind parameters for a quiet case, which we adopted in the present study, were the following:
\begin{itemize}
	\item Stellar wind density $n_{sw} = 2.5 \times 10^8$~m$^{-3}$;
	\item Stellar wind velocity $v_{sw} = 330$~km/s;
	\item Stellar wind temperature $T_{sw} = 3.5 \times 10^5$~K. 
\end{itemize}

Here and below, the stellar wind densities are the proton densities. Assumed stellar wind parameters correspond to a total mass loss rate of $3.5 \times 10^{-14}$~M$_\odot$~year$^{-1}$, or $\approx 2.5$ times the mass loss rate of the current Sun. This value is higher than the mass loss rate of GJ~436 estimated from the Ly$\alpha$ observations of GJ~436b by \citet{Vidotto17} of $(0.45 - 2.5) \times 10^{-15}$~M$_\odot$~year$^{-1}$. A higher mass loss rate leads to \textit{i)} a higher electron impact ionization rate and \textit{ii)} a higher amount of ENAs which are produced as a result of interaction between the atmospheric neutrals and incoming stellar wind. In Section~\ref{sec_wind}, we investigated the influence of different stellar wind parameters on our results.

A lower electron impact ionization rate would decrease the ionization of neutral hydrogen already present in the atmosphere and, therefore, slightly increase the modeled absorption for hydrogen-dominated atmospheres. We used the database of energy-dependent absorption cross sections for electron impact ionization by the National Institute of Standards and Technology (NIST\footnote{\url{https://physics.nist.gov/PhysRefData/Ionization/intro.html}}). We calculated the electron impact ionization rates used in the simulations as $\tau_{\rm ei} = \sigma(\epsilon) v_{\rm th} n_{\rm sw}$, where $\sigma(\epsilon)$ is the cross section for a given energy $\epsilon$ of a stellar wind, which is calculated according to its temperature as $\epsilon = 3/2 k_B T_{\rm sw}$, and $v_{\rm th} = \sqrt{3 k_B T_{\rm sw}/m_{\rm el}}$ is the most probable Maxwellian velocity of a electron. Here, $T_{\rm sw}$ is the stellar wind temperature. Electron impact ionization affects only hydrogen atoms outside the magnetospheric-ionospheric boundary, while photoionization affects all atoms exposed to sunlight (not in the shadow of the planet).

\subsection{Photoionization rates and the Ly$\alpha$ absorption rates}
\label{subsec_Petra}

To calculate the photoionization rates we used the XUV spectrum of GJ~436 from the MUSCLES database\footnote{\url{https://archive.stsci.edu/prepds/muscles}} \citep{France16,Youngblood16,Loyd16}. Specifically, we used the panchromatic spectral energy distribution binned to a constant 1\,\AA\ resolution, version 1.2. The X-ray ($<$100\,\AA) part was derived from \textit{Chandra} observations and the (unobservable) EUV (100-912\,\AA) part was reconstructed with the method of \citet{Linsky14} from the star's Ly$\alpha$ emission \citep{Loyd16}. The spectra contain the apparent fluxes as observed at the Earth, so we adopted a distance of 10.14\,pc (parallax of 98.61\,mas; \citealp{vanLeeuwen07}) and an orbital distance of 0.24\,AU to scale the fluxes to the center of GJ~436's HZ. For the photoionization cross sections $\sigma_\mathrm{ph}$ we used the analytic fits from \citet{Verner96}. We then calculated the photoionization rates $\beta_\mathrm{ph} = \int_{\lambda_0}^{\lambda_1} \sigma_\mathrm{ph} \phi_\mathrm{XUV} d\lambda$, where $\phi_\mathrm{XUV}$ is the XUV photon flux at 0.24\,AU, $\lambda_0=5$\,\AA\ (minimum wavelength of the adopted spectrum), and $\lambda_1$ is the wavelength of the ionization threshold of the specific atom. The resulting photoionization rates were $1.434\times10^{-7}$\,s$^{-1}$ for H and $1.158\times10^{-6}$\,s$^{-1}$ for N. The adopted spectrum yielded a total XUV (5-912\,\AA) flux of 15.6\,erg\,cm$^2$\,s$^{-1}$ at 0.24\,AU from the star, which is about three times the present solar value at Earth (4.64\,erg\,cm$^2$\,s$^{-1}$; \citealp{Ribas05}). The total XUV flux from MUSCLES is slightly lower than the value estimated by Ehrenreich et al. (2015), who derived a 35\% higher XUV flux. More recently, \citet{King18} estimated the XUV emission of GJ~436 to be 26\% higher than from MUSCLES. Since the X-ray range does not contribute significantly to the photoionization rates because of the small cross sections, the rates would be raised by 20-28\% if using these alternative XUV flux reconstructions. GJ~436 is variable in X-rays and likely also in EUV \citep{King18}, so data taken at different epochs may yield varying XUV flux estimates and thus photoionization rates. The photoionization rates at 0.24~AU for a more active GJ~436-like star were calculated as follows. For a given Ly$\alpha$ flux, the XUV spectrum can be calculated following \citet{Linsky13,Linsky14}. \citet{Linsky13} provide a scaling for the X-ray band ($<$100\AA) and \citet{Linsky14} for the EUV (100-912\AA). For properly scaled Ly$\alpha$ fluxes, we computed the corresponding XUV fluxes by summing X-ray and EUV contributions to obtain 7, 10 and 20 XUV. The spectra matching these total XUV fluxes were then used to obtain the photoionization rates (Table 2) with cross-sections taken from \citet{Verner96} and integrating over wavelength.

\begin{figure}
\includegraphics[width=1.0\columnwidth]{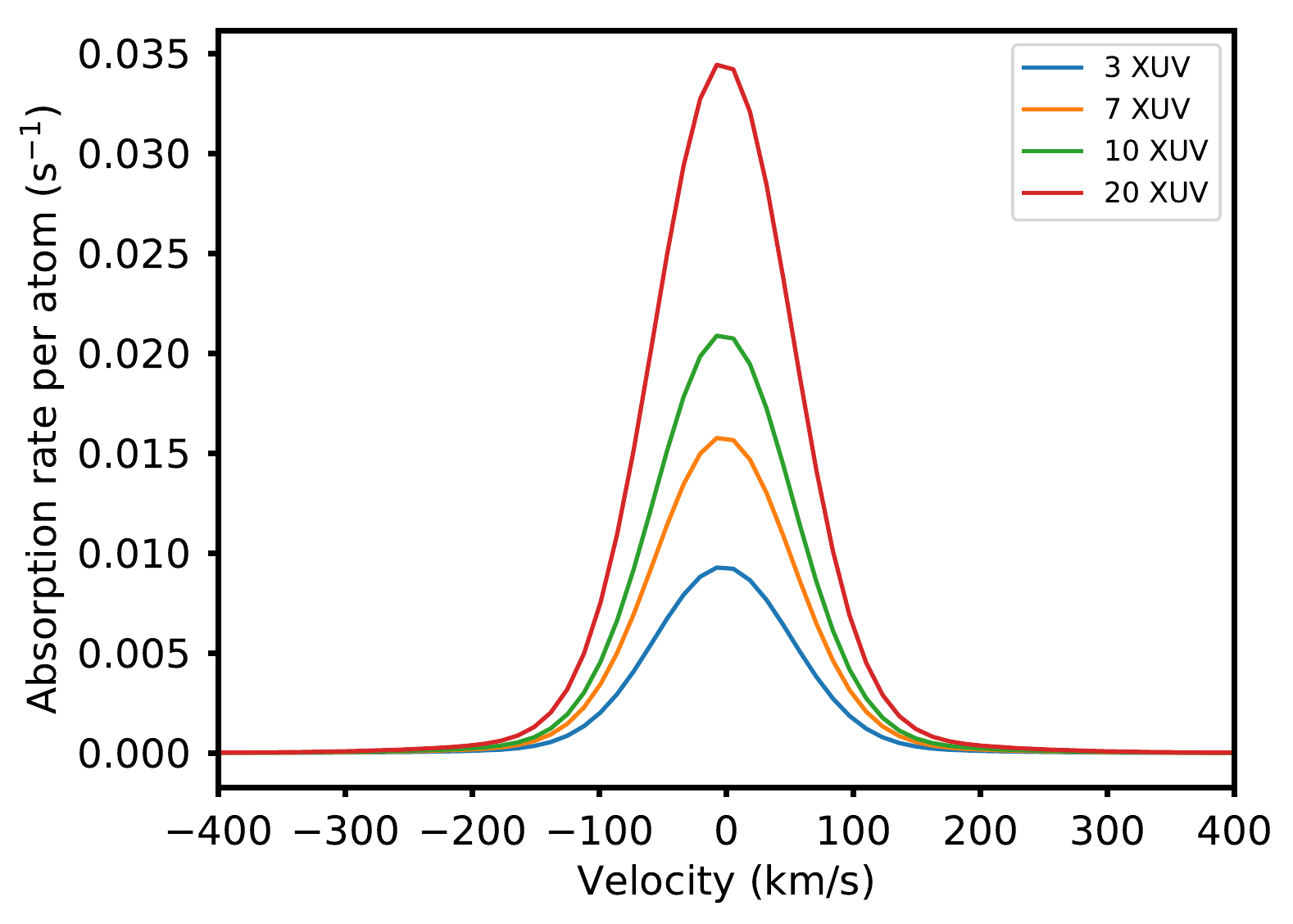}
 \caption{Ly$\alpha$ absorption rate dependent on the radial velocity of hydrogen atoms. Positive velocities denote motion toward the star.}
  \label{f_LyaGJ436}
\end{figure}

For obtaining the velocity-dependent Ly$\alpha$ absorption rate per atom $\beta_\mathrm{abs}(v) = \sigma_\mathrm{abs} \phi_\mathrm{Ly\alpha}(v)$ we used the reconstructed intrinsic line profile from the MUSCLES database \citep{Youngblood16}. Here, $\phi_\mathrm{Ly\alpha}(v)$ is the Ly$\alpha$ photon flux at 0.24\,AU, $\sigma_\mathrm{abs}=5.47\times10^{-15}$\,cm$^2$\,\AA$^{-1}$ the total Ly$\alpha$ absorption cross section \citep{Quemerais06}, and $v=-c(\lambda/\lambda_\mathrm{c}-1)$ is the atom's radial velocity (counted positive toward the star; $\lambda_\mathrm{c}=1215.67$\,\AA). The maximum absorption rate at the line peak is $9.2\times10^{-3}$\,s$^{-1}$, falling off by a factor of ten within $\pm$120\,km\,s$^{-1}$ from the line center. The total intrinsic Ly$\alpha$ flux from MUSCLES is in good agreement with \citet{B15}, but about 30-60\% lower than the reconstructions shown in earlier studies \citep{Ehrenreich11,France16}. These differences can be attributed to slightly different reconstruction methods, as well as using data taken at different epochs. Calculated absorption rates for 3, 7, 10, and 20 XUV are shown in Fig.~\ref{f_LyaGJ436}. To determine the UV absorption rates for XUV enhancement factors of 7, 10, and 20 we used the corresponding Ly$\alpha$ fluxes matching these enhancements from the method described above. We then scaled the present line profile of GJ 436 to these enhanced flux levels to obtain the corresponding UV absorption rates for the higher activity levels.

We did not consider acceleration by the radiation pressure for species other than H, because the stellar flux in the corresponding lines is much smaller and, therefore, species other than H are not accelerated efficiently. In this paper, we did not include self-shielding, meaning that we did not account for the optical depth of the hydrogen envelope in the Ly$\alpha$ line and assumed that all hydrogen atoms at all locations in the simulation domain (except in the optical shadow of the planet) have the same probability to scatter a stellar Ly$\alpha$ photon. This seems reasonable because in this paper we focused mostly on thin hydrogen envelopes (especially for nitrogen-dominated atmospheres). For hydrogen-dominated atmospheres, absence of self-shielding leads to excess acceleration of atmospheric particles away from the star. However, due to low radiation pressure in the habitable zone of GJ~436, this effect is rather minor and not so significant as for GJ~436b (see Fig.~\ref{f_LyaGJ436}).

\subsection{DSMC code}

In this subsection, we describe the DSMC upper atmosphere-exosphere 3D particle model used to study the plasma interactions between the stellar wind of GJ~436 and the upper atmospheres of terrestrial planets. The software is based on the FLASH code written in Fortran 90 \citep{Fryxell00}. The code follows particles in the simulation domain according to all forces acting on a hydrogen atom and includes up to three neutral species as well as their ions, with two of the species being neutral hydrogen and hydrogen ions (the latter also include stellar wind protons). 

The comprehensive description of the model can be found in \cite{H08,K14b}. The code has been successfully applied to modeling of hydrogen-dominated extrasolar giant \citep{K14b} and terrestrial planets \citep{K13,K14,Lichtenegger16}. The latest improvement of the code includes adding multiple species to it, which are allowed to charge exchange with stellar wind protons as well as being photo- and electron impact ionized and collide with each other. The main processes and forces included for an exospheric atom are:
\begin{enumerate}
 \item For hydrogen atoms: collision with a UV Ly-$\alpha$ photon which defines the velocity-dependent radiation pressure. 
 \item Charge exchange with a stellar wind proton.
 \item Elastic collision with another neutral atom.
 \item Ionization by stellar photons or wind electrons. 
 \item Gravity of the star and planet (including the tidal forces), centrifugal, and Coriolis forces.
\end{enumerate}

Summarizing this, our code includes three sources of ionization for atmospheric neutrals: photoionization, electron impact ionization, and charge exchange. Electron impact ionization and charge exchange can occur only outside the magnetospheric-ionospheric boundary, in the region, where the atmospheric neutrals come in contact and can collide with the stellar wind. While electron impact ionization is very important for hot Jupiters, in the HZ it is less significant due to a much less denser wind \citep{Holzer77}. If a planetary neutral is ionized, it is deleted from the simulation if it has been ionized inside the ionosphere and kept in the simulation, if the parental neutral particle has been ionized outside it. We assumed that all ions outside the ionosphere were picked up by the stellar wind and lost from the planetary atmosphere, thus contributing to ion pick up losses. The ions inside the ionosphere were deleted to shorten the simulation times. Our model does not include magnetic fields and collisions for the ions, therefore, the motion of the ions is governed only by the gravity forces of the planet and the star.
If a simulation does not include atmospheric neutral hydrogen, it is produced only as a product of charge exchange between the incoming stellar wind protons and atmospheric neutral particles. 

The magneto-ionopause of the planet was represented by a conic shaped obstacle:
\begin{equation}
  \label{e_obs}
  x = R_{\rm s} \left(1-\frac{y^2 + z^2}{R_{\rm t}^2} \right).
\end{equation}
Here, $R_{\rm s}$ stands for the planetary obstacle stand-off distance and $R_{\rm t}$ for the width of the obstacle. We used a Cartesian coordinate system with the $x$-axis pointing toward the star (meaning that particles flying toward the star have positive velocities along the $x$-axis, and particles flying away from the star have negative velocities), $y$-axis pointing opposite to the planetary motion, and, finally, $z$-axis completing the right-hand coordinate system. Since the obstacle shape and location depend strongly on the planetary magnetic field strength, one can model the interaction of the stellar wind with magnetized as well as with non- or weakly magnetized planets by the appropriate choice of $R_{\rm s}$ and $R_{\rm t}$. In this work, we were interested to estimate the maximum possible absorption due to production of ENAs, tus we selected the values of $R_{\rm s}$ and $R_{\rm t}$ just above the exobase, which maximized the size of the region where the stellar wind protons could penetrate in the atmosphere and interact with it. The height of the exobase was determined based on results of MHD modeling. This choice is also justified by the fact that the pressure balance point between stellar wind ram pressure and the thermal pressure of the planetary atmosphere (not accounting for the RAM pressure of the planetary outflow and magnetic pressure, since we do not consider magnetic fields in this study) was located below the inner boundary.

The obstacle was rotated by an angle of $\arctan (v_{\rm pl}/v_{\rm sw})$, to account for the finite stellar wind speed relative to the planet's orbital speed. This approach does not include polar openings (cusps), and as such can not reproduce all effects of a magnetosphere. However, it is a good choice for modeling of non-magnetized planets, on which we focused in this study. We assumed that the planets we consider lack intrinsic magnetic fields and experience a Venus-type interaction with the incoming stellar wind. In this case, the stellar wind plasma can directly impact the stellar atmosphere, leading to formation of a compressed and narrow ionospheric obstacle \citep{Lundin11}.

  
The code allows to track metaparticles of different weights. If charge exchange occurs between a stellar wind proton metaparticle and an atmospheric neutral particle of a different weight, then the bigger particle is split in two parts, one of which contains exactly the same amount of real atoms as the smaller metaparticle. The particle containing the excess atoms is excluded from the interaction and continues traveling with the same speed and trajectory. The particles with equal weights then charge exchange. In this paper, we include metaparticles of several different types (or species): neutral atomic hydrogen and nitrogen, ionized nitrogen of planetary origin, and two types of metaparticles for ionized hydrogen: of stellar and planetary origin.

After we modeled the hydrogen corona, we calculated the Ly$\alpha$ in-transit attenuation. Natural broadening was accounted for at this stage. Doppler broadening is included automatically by accurately taking into account the velocities of the atoms.

\subsection{Calculation of the Ly$\alpha$ transmissivity}
\label{sec_trans}

After a hydrogen cloud was simulated and by knowing the positions and velocities of all the hydrogen metaparticles at a certain time, we computed how these atoms attenuate the stellar Ly$\alpha$ radiation by using a post-processing software written in the Python programming language. To compute the transmissivity along the line-of-sight (LOS) we followed the approach of \citet{Semelin07}. The post-processing tools have been described in details in \citet{K14b}. Here we repeat the main features of them. 

To increase the particle statistics, especially for cases with a low number of metaparticles of a specific specie, particles were deposited in the cell according to the Cloud in cell (CIC) algorithm, where each particle is viewed as a cloud of particles with the same size as the cell. CIC was only applied for the spatial coordinates $(y, z)$. Each particle weight was distributed among four cells, which provided a self consistent smoothing of the solution. 

We assumed that the test planet has zero inclination, therefore, it was located in the center of the stellar disk at mid transit. However, we took inclination into account for GJ~436b, which is a grazing planet and is transiting very close to the edge of the stellar disk. Only neutral hydrogen atoms absorb in the Ly$\alpha$ line. One has to take into account spectral line broadening. Real spectral lines are subject to several broadening mechanisms: \textit{i)} natural broadening; \textit{ii)} collisional broadening; \textit{iii)} Doppler or thermal broadening. The ``natural line width'' is a result of quantum effects and arises due to the finite lifetime of an atom in a definite energy state. A photon emitted in a transition from this level to the ground state will have a range of possible frequencies: $\Delta f \sim \Delta E / \hbar \sim 1 / \Delta t $. The distribution of frequencies can be approximated by a Lorentzian profile. 
Collisional broadening is caused by the collisions randomizing the phase of the emitted radiation. This effect can become very important in a dense environment, yet above the exobase it does not play a role and is important only in the lower parts of the atmosphere.
The third type of broadening, which plays a significant role in the upper atmosphere of a hot exoplanet, is thermal broadening. If $f_0$ is the centroid frequency of the absorption line, the frequency will be shifted due to the Doppler effect. Combining Doppler shift with the Maxwellian distribution of $v_x$, one can obtain a Gaussian profile function, which is decreasing very rapidly away from the line center. The combination of thermal and natural (or collisional) broadening is described by the Voigt profile, which is the convolution of the Lorentz and Doppler profiles. 

In the considered cases, an analytical solution for the absorption profile cannot be obtained, since it is not only thermal atoms that contribute to the broadening. The presence of a non-thermal population of hot atoms (ENAs and atoms accelerated by the radiation pressure) changes the picture. Mathematically it means that the line width cannot be described by the Voigt profile anymore. We calculated the natural broadening for all atoms and bin it by velocity, which automatically gives us the Doppler broadening for a particular velocity distribution.

The velocity spectrum of an atomic cloud was then converted to frequency via the relation $f = f_0 + v_x / \lambda_0$ with $f_0 = c / \lambda_0$, $\lambda_0 = 1215.65 \times 10^{-10}$~m. To compute the transmissivity along the line-of-sight (LOS) we followed the approach of \citet{Semelin07}, where the relation between the observed intensity $I$ and the source intensity $I_0$ as a function of frequency $f$ could be written as
\begin{equation}
        T=I/I_0=e^{-\tau(f)}=e^{-\sigma(f)Q}.
        \label{e_T}
\end{equation}
Here, $\tau=\sigma(f)Q$ is the frequency dependent optical depth, $Q$ is the column number density of hydrogen atoms and $\sigma(f)$ is the frequency dependent crossection, which depends on the normalized velocity spectrum, the Ly$\alpha$ resonance wavelength and the natural absorption crossection in the rest frame of the scattered hydrogen atom \citep{Peebles93}. The quantity $T$ is called the transmissivity.
For a hydrogen cloud in front of a star, the transmissivity of the stellar spectrum was computed on an $yz$-plane which was discretized into a grid with $N_c$ cells. For each cell in the grid along lines of sight in front of the star ($y^2 + z^2 < R_{\rm st}^2$), we calculated the velocity spectrum of all hydrogen atoms in the column along the $x$-axis. Then we calculated the transmissivity averaged over all columns in the $yz$-grid except those particles which fell outside the projected limb of the star or inside the planetary disk. The average transmissivity was then calculated as 
\begin{equation}
        \bar{T}(f) = \frac{1}{N_c} \displaystyle\sum_{i=1}^{N_c} T_i(f),
        \label{e_avT}
\end{equation}
where $T_i$ is the transmissivity in a cell. For lines-of-sight in front of the planet $(y - y_p )^2 + (z - z_p )^2 < R_{\rm pl}^2$, where ($y_p$ , $z_p$) is the planet center position, we set $T_i = 0$ (zero transmissivity). Then, we appied the average transmissivity to the observed out-of-transit spectrum $I_0$ yielding the modeled in-transit spectrum
\begin{equation}
        I_m (f ) = I_0 (f )\bar{T}(f).
        \label{e_mSp}
\end{equation}
The frequency dependent crossection was defined by:
\begin{equation}
                \sigma(f)=\int_{-\infty}^{+\infty}\check{u}(v_x)~\sigma_N[(1+v_x/c)f]dv~[m^2],
\end{equation}
where $\check{u}(v_x)$ is the normalized velocity spectrum along the LOS, so that $\int_{-\infty}^{+\infty}\check{u}(v_x)dv_x=1$, $c$ is the speed of light.
We assumed the natural absorption crossection in the rest frame of the scattered hydrogen atom according to \cite{Peebles93}
\begin{equation}
        \sigma_N(f)=\frac{3 \lambda_{\alpha}^2 A^2_{21}}{8\pi} \frac{(f/f_{\alpha})^4}{4\pi^2(f-f_{\alpha})^2+(A_{21}^2/4)(f/f_{\alpha})^6}~[m^2].
        \label{e_crossec}
\end{equation}
Here, $f_{\alpha}=c/\lambda_{\alpha}$ with $\lambda_{\alpha}$ being the Ly$\alpha$ resonant wavelength and $A_{21}=6.265 \times 10^8$~s$^{-1}$ the rate of radiative decay from the $2p$ to the $1s$ energy level.
Multiplying $\sigma(f)$ with Q, we computed the optical depth directly without normalizing the velocity spectrum:
\begin{equation}
        \tau(f)=\int_{-\infty}^{+\infty} u(v_x)~\sigma_N[(1+v_x/c)f]dv.
        \label{e_tau}
\end{equation}
The expression for the natural absorption crossection given in Eq.~\ref{e_crossec} was approximated by the Lorentzian profile
\begin{equation}
        \sigma_N(f)=f_{12} \frac{\pi e^2}{m_e c} \frac{\Delta f_L/2\pi}{(f-f_{\alpha})^2 + (\Delta f_L /2)^2},
        \label{e_crossLor}
\end{equation}
where $f_{12}= 0.4162$ is the Ly$\alpha$ oscillator strength, $e$ is the elementary charge, $m_e$ is the electron mass, and $\Delta f_L = 9.936 \times 10^7$~[s$^{-1}$] is the natural line width \citep{Semelin07}. We used the crossection from Eq.~\ref{e_crossLor} to include the contribution of broadening to the absorption.
To account for the contribution of the lower atmosphere, for a case of a hydrogen-dominated atmosphere we added a Maxwellian velocity spectrum corresponding to a hydrogen gas with a specified column density and temperature to all pixels inside the inner boundary $R_{\rm ib}$. 

\section{Model validation. GJ~436b.}
\label{sec_GJ436b}

In this section, we validate our model by reproducing the Ly$\alpha$ transit observations of a ``warm Neptune'' exoplanet GJ~436b. Ly$\alpha$ transit observations by the Hubble Space Telescope revealed a strong absorption of about 8\% at mid-transit and 33\% post-transit \citep{Kulow14}. A later result which used a corrected ephemeris revealed an even deeper transit depth of $\approx$56\% in the Ly$\alpha$ line \citep{Ehrenreich15}, with ultraviolet transits repeatedly starting about two hours before, and end more than three hours after the approximately one hour optical transit. For comparison, in the optical wavelengths the transit depth is only about 0.69 \%. Later observations \citep{Lavie17} confirmed the presence of a deep in-transit absorption, and also showed that the Ly$\alpha$ transit started about two hours before the optical transit, and lasted for more than 10 hours after the end of the optical transit. Such deep absorption in the Ly$\alpha$ line is observed even despite GJ~436b grazing very close to the edge of the stellar disk, thus indicating a presence of a very extended hydrogen envelope enwrapping the planet. The time evolution of the observed Ly$\alpha$ emission line of GJ~436 before, during, and after the planetary transit is rather complex \citep{Lavie17}; previously, significant efforts have been undertaken to model the observed phenomena \citep{B15,B16}. Here, we prove that our model was also capable of reproducing the observed mid-transit absorption of GJ~436b. 

\begin{figure}
\includegraphics[width=0.9\columnwidth]{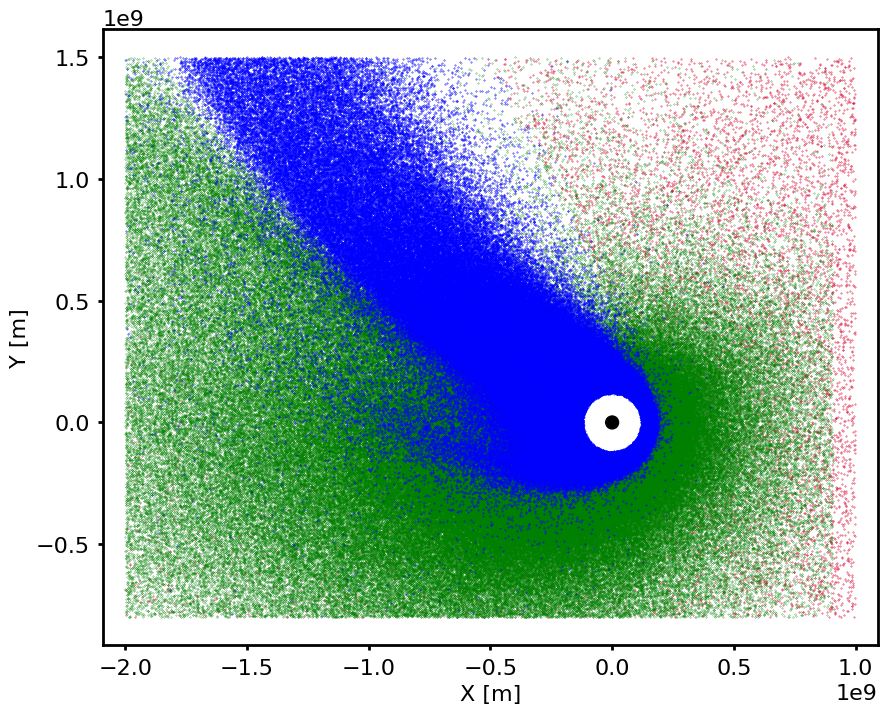}
 \caption{Atomic corona surrounding GJ~436b simulated with the DSMC code. Only particles with coordinates $-10^8 \le z \le 10^8$~m are shown. The black dot in the center shows the planet, the white area around the planet is the inner atmosphere modeled with a hydrodynamic code. The star is on the right. Blue dots are the neutral hydrogen atoms, which contribute to the Ly$\alpha$ absorption. Red dots are the stellar wind protons, and green dots are the ionized hydrogen atoms of planetary origin. Radiation pressure is included. }
  \label{f_GJ436b_cloud}
\end{figure}

\begin{table}
 \caption{Stellar, planetary, and stellar wind parameters used in our test simulation for GJ~436b. The stellar and planetary data are from \url{www.exoplanet.eu} and \citet{B16}, stellar wind parameters are adopted from \citet{Vidotto17}, except of a slightly faster stellar wind speed of 110~km/s instead of 85~km/s. Simulation inner boundary for the DSMC code has been adopted close to the Roche lobe $@ 5.83$~R$_{\rm pl}$. The parameters at the inner boundary have been calculated by the hydrodynamic code. The shift of the planet relative to the stellar center in the direction of the $z$ axis has been calculated according to the inclination. Assumed collision cross sections can be found in Table~\ref{t1}.  }
 \begin{tabular}{@{}lc}
  \hline
  Parameter & Value \\
  \hline
  Star mass $M_{\rm st}$ [kg] &  $8.98\times10^{29}$   \\
  Star radius $R_{\rm st}$ [m]  & $3.22\times10^{8}$ \\
  Planet's mass $M_{\rm pl}$ [kg] &  $1.32\times10^{25}$   \\
  Planet's radius $R_{\rm pl}$ [m]  & $2.65\times10^{7}$ \\
  Semi-major axis $d_{\rm HZ}$ [AU]  & 0.0288  \\ 
  Inclination [degree]  &  85.8 \\  
  SW number density $n_{\rm sw}$ [m$^{-3}$] &  $2.0 \times 10^9$  \\
  SW velocity $v_{\rm sw}$ [m$\cdot$s$^{-1}$] & $110\times10^{3}$  \\ 
  SW temperature $T_{\rm sw}$ [K] & $0.41\times10^{6}$   \\
  Inner boundary radius [m]  & $1.54\times10^{8}$ \\
  Inner boundary density [m$^{-3}$]  & $2.5\times10^{12}$ \\  
  Inner boundary temperature [K]  & 2693 \\    
  Electron impact ionizaion rate (hydrogen) $\tau_{\rm ei}$ [s$^{-1}$]   & $5.6\times10^{-5}$ \\
  Photoionizaion rate (hydrogen) $\tau_{\rm ph}$ [s$^{-1}$]   & $8.8\times10^{-6}$ \\
  Bulk atmospheric outflow velocity [m~s$^{-1}$]  & $7.9\times10^{3}$ \\      
  Magnetospheric obstacle substellar point $R_{\rm s}$ [m]  & $1.6\times10^{8}$ \\  
  Magnetospheric obstacle width $R_{\rm t}$ [m]  & $1.6\times10^{8}$ \\    

  \hline
 \end{tabular}
 \label{t0}
\end{table}

\begin{figure}[h]
\includegraphics[width=1.0\columnwidth]{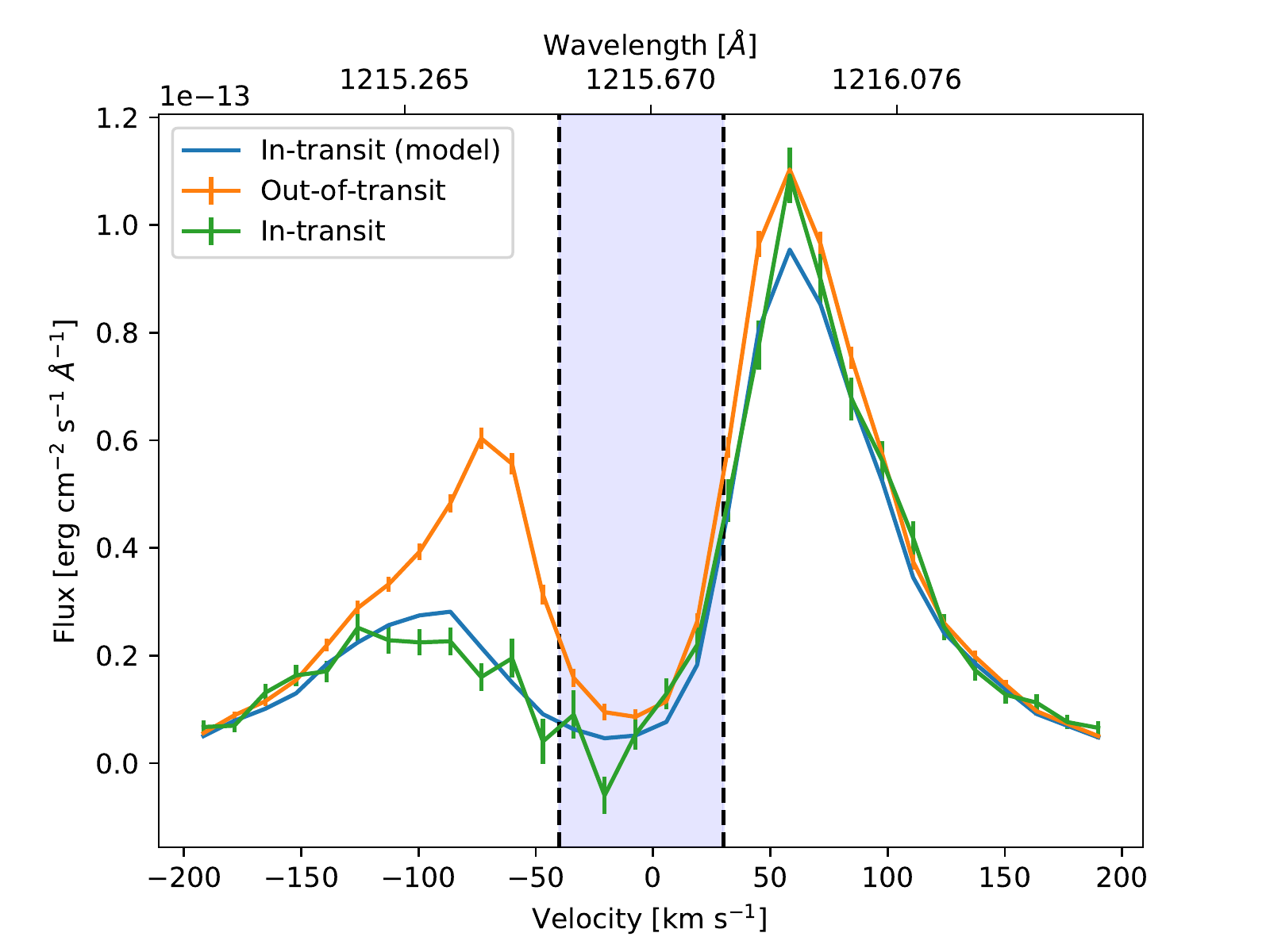}
 \caption{Modeled and observed mid-transit Ly$\alpha$ absorption of GJ~436b. We have used the in-transit data by \citet{Lavie17} for out-of-transit and mid-transit observations. The core of the Ly$\alpha$ line has been excluded, because it has been contaminated by geocoronal emission and thus cannot be observed. The contaminated region is marked by the dashed vertical lines and the blue shading. The green and orange lines show the mis-transit and out-of-transit observations (\citealp{Lavie17}, Fig.~3). The blue line shows the modeled in-transit absorption. } 
  \label{f_GJ436b_absorption}
\end{figure}

We applied the model described in the previous section to GJ~436b. First, we model the lower atmosphere using the hydrodynamic code described in Section~\ref{sec_atmpar}. To calculate the lower atmosphere velocity, density, and temperature profiles, we assumed the lower atmosphere of GJ~436b to be comprised of atomic hydrogen only. Close to the planetary radius, the main atmospheric constituent is molecular H$_2$, which we did not take into account. However, sophisticated 3D modeling shows that the H$_2$ dissociation front is reached already at 1.35$R_{\rm pl}$, therefore, atomic hydrogen is the main atmospheric component thorough most of our simulation domain for the lower atmosphere \citep{Shaikhislamov18}. Some works \citep{Hu15} have shown that planets like GJ~436b might accumulate a significant amount of helium due to faster atmospheric escape of hydrogen. The presence of helium can significantly influence the mass loss rate and the atmospheric structure \citep{Shaikhislamov18}. However, although our hydrodynamic model for the lower atmosphere is 1D and not 3D and although we did not take into account the heavy species in this modeling, the model provided very similar values for hydrogen number density, temperature, and outflow velocity at 5.83~R$_{\rm pl}$ in comparison to the 3D multiple-species model \citep{Shaikhislamov18}. We assumed the stellar wind parameters at GJ~436b's orbit as estimated by \citet{Vidotto17} except of a slightly faster stellar wind, which, however, still fell well inside the range of the mass loss rates predicted for GJ~436 by \citep{Vidotto17}.

Our results confirmed that GJ~436b possesses an extended hydrogen envelope, the presence of which can be anticipated from the deep Ly$\alpha$ in-transit absorption signature and has been confirmed by previous modeling \citep{B16}. Due to relatively low gravity of this planet, the exobase is located higher than the Roche lobe, above which the atmosphere is not bound to the planet gravitationally. This means that the atmosphere stays collisional up until the Roche lobe and beyond. According to our simulations, the Roche lobe and the exobase were located at 5.83~R$_{\rm pl}$ and 7.67~R$_{\rm pl}$, respectively. We started our DSMC simulation at the Roche lobe of the planet, assuming the density, temperature, and outflow velocity of the atomic hydrogen gas obtained in the hydrodynamic simulation (described in Section~\ref{sec_atmpar}) as an input at the spherical inner boundary. Table~\ref{t0} summarizes simulation input parameters for the DSMC simulation. As in all other simulations, we took into account electron impact and photoionization, elastic collisions between the atmospheric particles, and charge exchange collisions with the stellar wind protons. We accounted for the radiation pressure at the orbital location of GJ~436b (Fig.~\ref{f_LyaGJ436}; the data for 3~XUV in the habitable zone can be scaled according to the closer orbital distance of GJ~436b), but we did not take into account the self-shielding, which means the optical depth of the atmosphere for the Ly$\alpha$ radiation \citep{B16}. Fig.~\ref{f_GJ436b_cloud} shows the modeled atomic corona around GJ~436b. Fig.~\ref{f_GJ436b_absorption} shows the calculated mid-transit Ly$\alpha$ absorption in comparison with observations. A test run without radiation pressure (not shown here) yielded similar result to the one shown in Fig.~\ref{f_GJ436b_absorption} with only slight overabsorption in the red wing of the Ly$\alpha$ line, which indicates that the atmosphere is escaping very efficiently due to gravitational forces. This seems reasonable, because due to the proximity to its host star, GJ~436b has a Roche Lobe located at only 5.83~R$_{\rm pl}$, with a significant fraction of the atmosphere being above it. After the particles leave the planet's gravitational well, they are no longer bound to the planet gravitationally and escape, forming a long tail curved by the Coriolis force (Fig.~\ref{f_GJ436b_cloud}). We could reproduce the observations assuming a lower velocity of the planetary wind in comparison to earlier works \citep{B16}, which assumed the outflow velocities of the planetary gas of 50--60~km~s$^{-1}$. In our model, we assumed an outflow velocity of 7.9~km~s$^{-1}$ at the inner boundary, which was the gas velocity at this distance to the planet in accordance to our hydrodynamic model applied to the lower atmosphere. We could obtain a good fit to the observations also assuming the stellar wind speed of 85~km/s as estimated by \citet{Vidotto17}, however, we got a slight overabsorption on the right part of the Ly$\alpha$ line in comparison to 110~km/s which is shown in Fig.~\ref{f_GJ436b_absorption}.

We have calculated the Ly$\alpha$ in-transit signature of GJ~436b at mid-transit and compared it to the observed mid-transit signature by \citet{Lavie17}. Our model agrees well with the observations both on the left and right wings of the Ly$\alpha$ line with a overabsorption in only one observational point on the right wing, therefore confirming that out model can be used to estimate the in-transit absorption depth in the Ly$\alpha$ line of exoplanets. Detailed modeling of the Ly$\alpha$ observations of GJ~436b will be published in a following article (currently in preparation).   

\section{Simulation parameters for terrestrial planets}
\label{sec_parameters}

In this section, we summarize our simulation parameters which we have used as inputs for the DSMC code. In all cases, the planet has been assumed to have a mass and radius equal to the Earth, and to orbit in the middle of the HZ of GJ~436 at 0.24~au. The location of the habitable zone has been calculated as in \citet{K13}. 

Stellar wind parameters have been chosen according to \citet{K13} and have been the same in all simulations, as described in Section~\ref{sec_sw}. Since we assumed a constant stellar wind, the ionization rate by stellar wind electrons is also constant in all simulations. Electron impact ionization rates have been caclulated according to \citet{BanksKocharts1973}. All simulation parameters which were kept constant for all XUV fluxes are summarized in Table~\ref{t1}. 

\begin{table}
 \caption{Stellar, planetary, and stellar wind parameters used in all simulations. Stellar wind and habitable zone parameters of GJ436 have been adopted from \citet{K13}, electron impact ionization rates were calculated using the NIST databases. Charge exchange cross sections are from \citet{Lindsay05} for H$^+$-H and \citet{Rinaldi11} for N$^+$-H. Elastic collision crossections were calculated as described in section~\ref{sec_atmpar}.  }
 \begin{tabular}{@{}lc}
  \hline
  Parameter & Value \\
  \hline
  Planet's mass $M_{pl}$ [kg] &  $5.97\times10^{24}$   \\
  Planet's radius $R_{pl}$ [m]  & $6.371\times10^{6}$ \\
  HZ distance $d_{\rm HZ}$ [AU]   & 0.12 -- 0.36  \\
  Orbital distance $d_{\rm orb}$, [AU]   & 0.24  \\
  Electron impact ionizaion rate (N) $\tau_{\rm ei}$ [s$^{-1}$]   & $6.3\times10^{-6}$ \\
  Electron impact ionizaion rate (H) $\tau_{\rm ei}$ [s$^{-1}$]   & $1.0\times10^{-5}$ \\
  SW number density $n_{\rm sw}$ [m$^{-3}$] &  $2.5 \times 10^8$  \\
  SW velocity $v_{\rm sw}$ [m$\cdot$s$^{-1}$] & $330\times10^{3}$   \\
  SW temperature $T_{\rm sw}$, [K] & $2\times10^{6}$   \\
  Charge exchange crossection H$^+$-H [m$^2$] & $2.0\times10^{-19}$   \\  
  Charge exchange crossection H$^+$-N [m$^2$] & $1.0\times10^{-19}$   \\
  H-H elastic collision crossection [m$^2$] & $3.5\times10^{-20}$   \\  
  H-N elastic collision crossection [m$^2$] & $3.7\times10^{-20}$   \\    
  H-N elastic collision crossection [m$^2$] & $3.9\times10^{-20}$   \\      
  \hline
 \end{tabular}
 \label{t1}
\end{table}

In this paper, we were interested in modeling of the strongest possible interaction between the stellar wind and the planetary atmosphere. For this reason, we focused on non-magnetized planets. If the planet has no intrinsic magnetic field, we can assume that the magnetospheric-ionospheric obstacle lies very close to the inner atmosphere boundary and is very narrow (similar to the ionospheric boundary observed at Venus; \citealp{Galli08,Lundin11}). In our model, including an intrinsic magnetic field reduces the size of the region where the stellar wind can interact with the planetary atmosphere, thus reducing the number of produced ENAs and the ENAs absorption in the wings of the Ly$\alpha$ line. We did not model the polar cusp regions where the stellar wind can interact with neutral atmospheres of magnetized planets. Around the Earth, ENAs have been observed both at the magnetopause region \citep{Fuselier10} and in the cusps \citep{Petrinec11}. Since the latter region of ENAs formation is typical only for a planet with an intrinsic magnetic field, we focus only at the interaction at the substellar region of the planetary magnetospheric-ionospheric boundary. Both the atmosphere and the ENAs contribute to the total in-transit absorption in the Ly$\alpha$ line. The atmosphere contribute mostly through the spectral line broadening (e.g., \citealp{BJ07}). Atmospheric particles can also be accelerated up to high velocities by the radiation pressure of the host star and increase the absorption in the left wing of the Ly$\alpha$ line. This mechanism is extremely effective for close-in planets such as HD~209458b or, to a lesser extent, GJ~436b (e.g., \citealp{BL13,B16}), but is of less importance for planets in the habitable zone, where the radiation pressure is naturally weaker due to the larger orbital distances (see Fig.~\ref{f_LyaGJ436}). In this paper, we did account for the atmosphere contribution via line broadening and for the radiation pressure. For a given radiation pressure and an atmospheric density profile, the maximum amount of ENAs is produced at a non-magnetized planet.

\begin{table*}
 \caption{Atmospheric parameters used in the simulations for nitrogen- and hydrogen-dominated atmospheres for 3, 7, 10, and 20~XUV. Adopted from \citet{Tian08} (3 and 20~XUV) and \citet{L08} (7 and 10~XUV). The value of the hydrogen density at the modern terrestrial exobase has been adopted from \citet{K13}. The value of $R_{\rm s}$ is chosen arbitrarily slightly above the exobase.}
 \begin{tabular}{@{}lcccc}
  \hline
  Parameter & 3~XUV & 7~XUV & 10~XUV & 20~XUV \\
  \hline
  Nitrogen-dominated atmosphere with Earth-like hydrogen content &  &   & &      \\
  $R_{\rm ib}$ [m]           & $7.4\times10^6$  & --  & --  & --     \\
  $n_{\rm ib}$ [m$^{-3}$] (N)& $1.2\times10^{13}$  &  &     \\
  $n_{\rm ib}$ [m$^{-3}$] (H)& $7.0\times10^{10}$  &  &     \\
  $R_{\rm s}$  [m]           & $7.4\times10^6$  &  &  &       \\
  $T_{\rm ib}$ [K]           & $2.2\times10^3$  &  &  &      \\
  \hline

  Nitrogen-dominated atmosphere &  &   & &      \\
  $R_{\rm ib}$ [m]           & $7.4\times10^6$     &  $2.15\times10^7$    & $2.7\times10^7$     & $8.6\times10^7$      \\
  $n_{\rm ib}$ [m$^{-3}$] (N)& $1.2\times10^{13}$  &  $2.2\times10^{12}$  & $4.6\times10^{11}$  & $	4.4\times10^{11}$  \\
  $R_{\rm s}$  [m]           & $7.4\times10^6$     &  $2.23\times10^7$    & $2.8\times10^7$     & $8.64\times10^7$      \\
  $T_{\rm ib}$ [K]           & $2.2\times10^3$     &  $7.5\times10^3$     & $5.0\times10^3$     & $2.53\times10^3$      \\
  \hline
  Hydrogen-dominated atmosphere   & &  & &      \\
  $R_{\rm ib}$ [m]           & $4.25\times10^7$  &  $5.65\times10^7$      & $6.3\times10^7$     & $8.2\times10^7$      \\
  $n_{\rm ib}$ [m$^{-3}$] (H)& $2.5\times10^{11}$  &  $1.74\times10^{11}$ & $1.54\times10^{11}$ & $1.07\times10^{11}$  \\
  $R_{\rm s}$  [m]           & $4.31\times10^7$  &  $5.73\times10^7$      & $6.37\times10^7$    & $8.28\times10^7$      \\
  $T_{\rm ib}$ [K]           & $1.14\times10^3$  &  $2.4\times10^3$       & $3.04\times10^3$    & $4.69\times10^3$      \\
  \hline
  Species photoionization rates   &   & &   &   \\
  H photoion. rate [s$^{-1}$] & $1.43\times10^{-7}$  & $3.17\times10^{-7}$  & $4.82\times10^{-7}$  & $1.03\times10^{-6}$  \\
  N photoion. rate [s$^{-1}$] & $1.15\times10^{-6}$  & $2.14\times10^{-6}$  & $2.97\times10^{-6}$  & $5.43\times10^{-6}$ \\
  \hline
 \end{tabular}
 \label{t2}

 \medskip
 Here, $R_{\rm ib}$ is the inner boundary radius, $T_{\rm ib}$ is the inner boundary temperature, $n_{\rm ib}$ is the inner boundary number density, $R_{\rm s}$ is the obstacle standoff distance. We assume the obstacle width equal to distance to the substellar point, therefore, $R_{\rm} s = R_{\rm t}$.
\end{table*}

Table~\ref{t2} summarizes our atmospheric parameters for all simulations. As one can see, a nitrogen-dominated atmosphere expands significantly when the XUV flux is increased from three to seven times the modern Earth level. The exobase temperature increases from 3 to 7~XUV and then drops again due to adiabatic cooling caused by atmospheric expansion. Hydrogen-dominated atmospheres also expand if the XUV flux is increased.  We always chose the inner boundary of our simulation, $R_{\rm ib}$, equal to the exobase height estimated in 1D simulations by \citet{L08} for nitrogen-dominated atmospheres and the 1D model described in section~\ref{sec_atmpar} adopted from \citet{Erkaev17} for hydrogen-dominated atmospheres.

\section{Results.}
\label{sec_results}

\subsection{Modeling of atomic coronae around Earth-like planets.}

Here, we present the results of our modeling of planetary exospheres of planets orbiting in the center of the habitable zone of GJ~436 (at 0.24~au). As stated earlier, we modeled on an Earth-like planet with nitrogen-dominated and hydrogen-dominated atmospheres at 3, 7, 10, and 20~XUV. At 3~XUV, we additionally model a nitrogen-dominated atmosphere with a terrestrial hydrogen content. 

For every atmosphere, the velocity distribution of neutral hydrogen is modified by the stellar radiation pressure, which accelerates the atoms in the direction away from the star according to Fig.~\ref{f_LyaGJ436} (however, in the habitable zone this effect is less impotant than for hot Jupiters or for a close-in planet like GJ~436b). We assume that neutral nitrogen atoms are not accelerated by the stellar radiation, as for them this effect is negligible. The distribution of neutral atoms (both hydrogen and nitrogen) is also altered by photo- and electron impact ionizations, which predominantly remove particles with positive velocities along the x-axis (flying toward the star), as they are more probable to leave the shadow of the planet and the ionosphere boundary and undergo ionization.

\begin{figure*}[h]
\includegraphics[width=2.0\columnwidth]{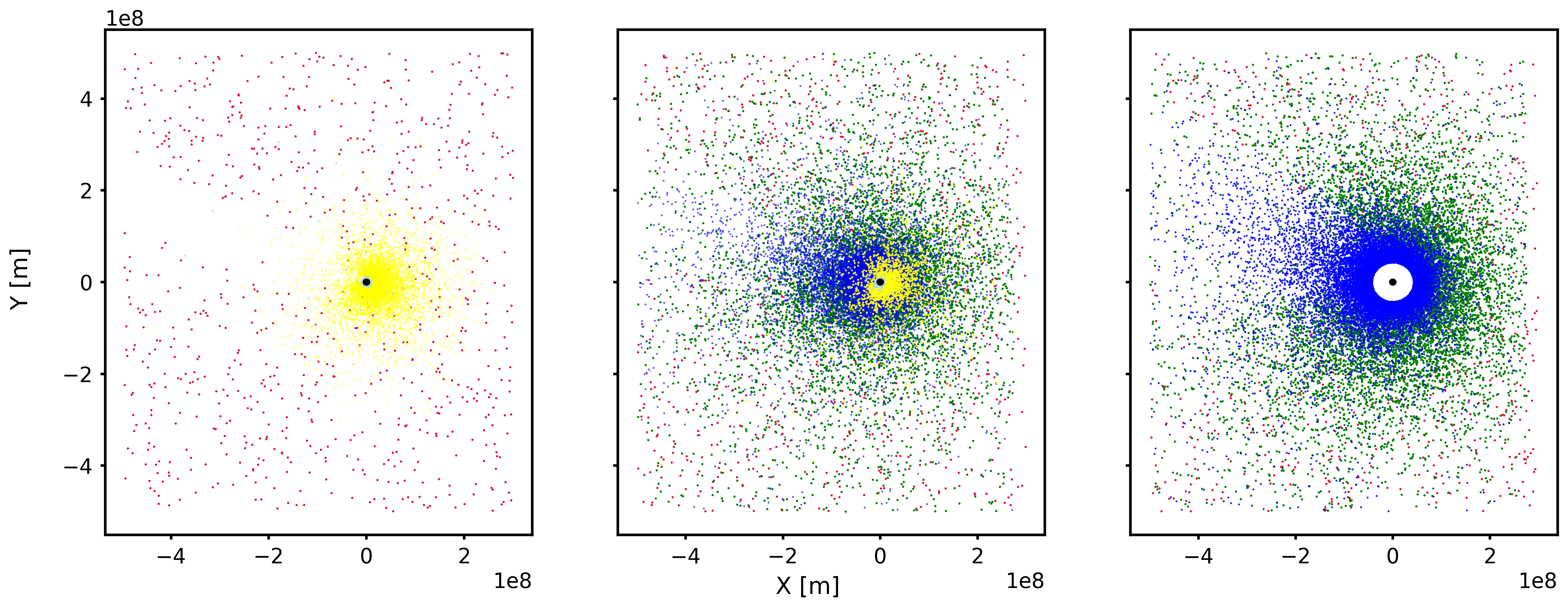}
 \caption{Illustration of modeled 3D atomic hydrogen coronae for 3 XUV (current flux in the habitable zone of GJ~436). Presented is a slice of modeled 3D domain for $-10^6<z<10^6$~m. Left: nitrogen-dominated atmosphere. Middle: nitrogen-dominated atmosphere with Earth-like hydrogen content. Right: hydrogen-dominated atmosphere. All dots show simulated metaparticles of the exosphere of a planet, and stellar wind. The red dots symbolize stellar wind protons, the blue dots are the neutral atmospheric hydrogen particles, the light blue dots are neutral nitrogen atoms and, finally, the yellow dots are ionized planetary N+. The green dots are ionized planetary hydrogen atoms (H+ of planetary origin). The black dot in the center of each plot symbolizes the planet, the white area around it is the lower atmosphere which is not included in the DSMC simulation. For nitrogen-dominated atmospheres, the lower atmosphere is very close to the planetary surface, so that the white area is not visible on the plots due to scaling. The star is on the right, and the stellar wind is coming from the right side. }
  \label{f_clouds3XUV}
\end{figure*}

\begin{figure*}
\includegraphics[width=2.0\columnwidth]{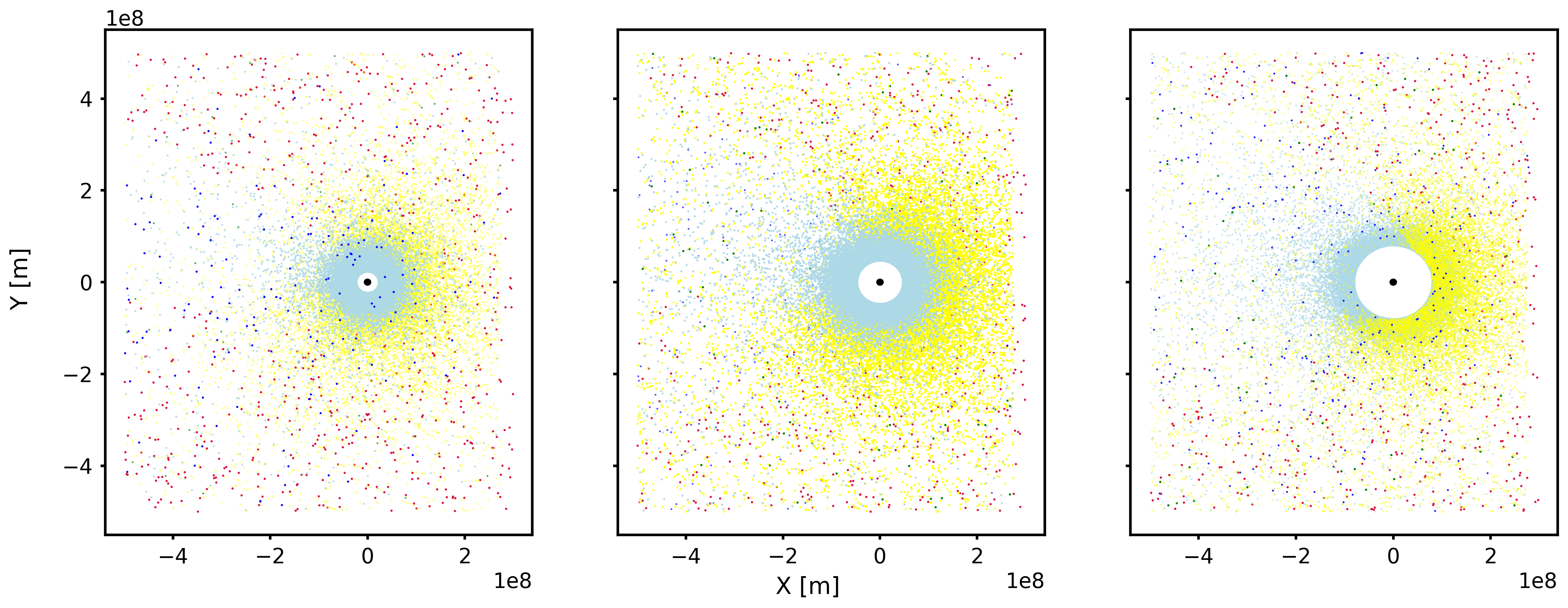}
 \caption{Same as Fig.~\ref{f_clouds3XUV}, but for a nitrogen-dominated atmosphere with no initial hydrogen content for 7, 10, and 20 XUV (from left to right). One can see that the inner atmosphere expands at high XUV fluxes. The yellow dots (N ions) indicate a region of enhanced ionization. The dark blue dots are the neutral hydrogen atoms produced due to charge exchange between neutral N atoms and incoming stellar wind protons, the light blue dots are neutral N atoms.}
  \label{f_cloudsNitrogen}
\end{figure*}

\begin{figure*}
\includegraphics[width=2.0\columnwidth]{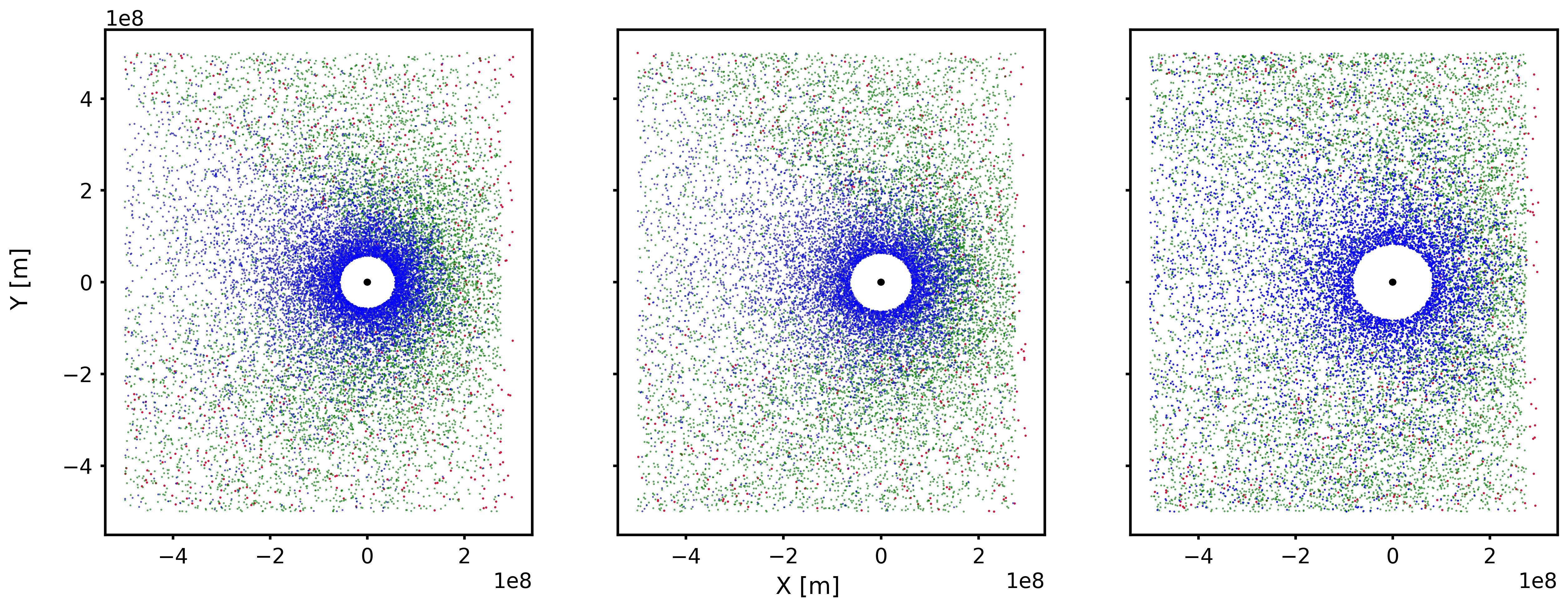}
 \caption{Same as Figs.~\ref{f_clouds3XUV} and \ref{f_cloudsNitrogen}, but for a a hydrogen-dominated atmosphere at 7, 10, and 20 XUV.}
  \label{f_cloudsHydrogen}
\end{figure*}

Figs.~\ref{f_clouds3XUV}, \ref{f_cloudsNitrogen}, and \ref{f_cloudsHydrogen} show the simulated exospheric atomic clouds around the planet. All figures show all particles within a slice with coordinates $-10^6 \le z \le 10^6 $~m. The ionospheric boundary is not explicitly shown, but can be seen as an area close and behind the planet devoid of stellar wind protons. 

 Fig.~\ref{f_clouds3XUV} illustrates the appearance of the atomic clouds for 3 XUV for three different atmospheric compositions: a nitrogen-dominated atmosphere, a nitrogen-dominated atmosphere with a modern Earth hydrogen fraction, and, finally, a hydrogen-dominated atmosphere. As one can see, the hydrogen atmosphere is very expanded even if exposed to this comparably low XUV~level. 
Fig.~\ref{f_cloudsNitrogen} shows three exospheric clouds simulated assuming a nitrogen-dominated composition exposed to 7, 10, and 20~XUV. As one can see, a pure nitrogen atmosphere exposed to XUV fluxes higher than $\sim$5--7 times the modern Earth value heats up efficiently, and, in absence of strong infrared coolants, the lower atmosphere expands to high altitudes, so that the exobase is located at very high altitudes in comparison to the 3~XUV case. In Fig.~\ref{f_cloudsNitrogen}, the lower atmosphere is shown as a white empty area surrounding the black dot, the planet. High exobase altitudes lead to intense charge exchange between the stellar wind and the atmosphere due to the increased planetary cross section, and the production of a much higher amount of ENAs (see velocity distribution of neutral hydrogen atoms in section~\ref{sec_spectrum}). Finally, Fig.~\ref{f_cloudsHydrogen} shows our modeling results for a hydrogen-dominated planet exposed to 7, 10 and 20~XUV. At a given XUV flux, a hydrogen-dominated atmosphere is even more expanded than a nitrogen-dominated one, which is due to the low mass of a hydrogen atom. Summarizing the results shown in Figs.~\ref{f_clouds3XUV}, \ref{f_cloudsNitrogen} and \ref{f_cloudsHydrogen}, we can conclude that both hydrogen and nitrogen-dominated atmospheres react with expansion to an enhanced level of stellar short wavelength radiation.

\subsection{Modeling of in-transit absorption in the Ly$\alpha$ line.}
\label{sec_spectrum}

In this section, we calculated the absorption in the Ly$\alpha$ line which would be produced if a planet surrounded by exospheric clouds shown in the previous section would transit in front of GJ~436. The method of calculating the Ly$\alpha$ transmissivity has been presented in section~\ref{sec_trans}.

First, we calculated the velocity distribution of all neutral hydrogen atoms which were present in the simulation domain at the end of the simulation. For a hydrogen-dominated atmosphere, hydrogen atoms can be divided into two big groups: neutral exospheric particles of planetary origin with the velocity distribution slightly modified by the radiation pressure and ionization, and energetic neutral atoms produced as a result of the charge exchange between stellar wind protons and neutral particles. ENAs have a different velocity distribution in comparison with the planetary particles, since they initially keep the velocity of a stellar wind particle. For this reason, ENAs are much more energetic than the atmospheric particles. As we show below, ENAs contribute to a high velocity particle population. In case of nitrogen-dominated atmospheres, ENAs are the only source of exospheric hydrogen. For the case of 3~XUV we additionally modeled an atmosphere with an Earth-like hydrogen fraction, which in case of the Earth is produced by water photodissociation. 

\begin{figure*}
\includegraphics[width=2.0\columnwidth]{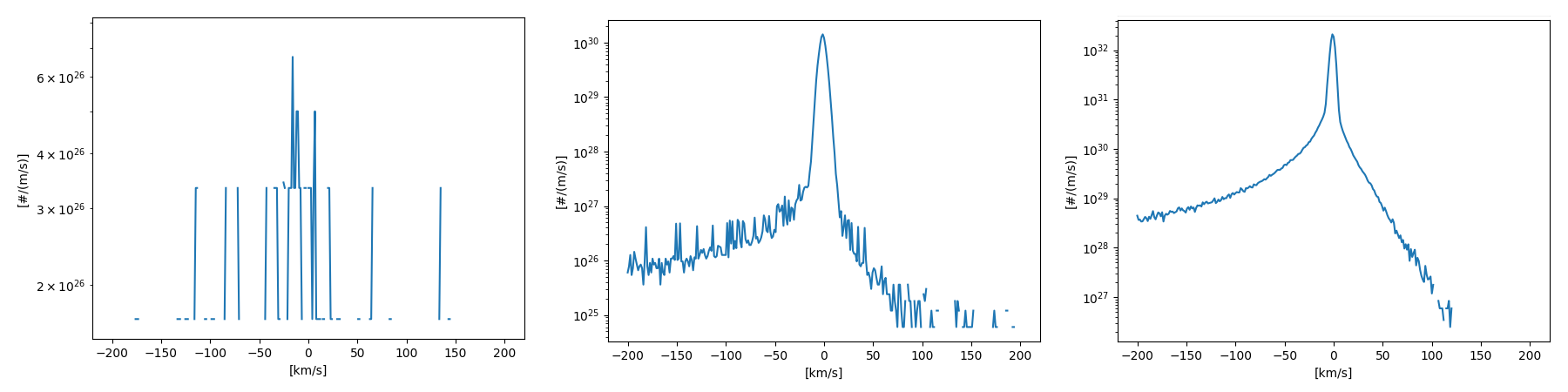}
 \caption{Velocity spectrum of neutral H atoms for different atmospheres exposed to the XUV flux 3 times the level at the Earth. From left to right: velocity spectrum for a nitrogen-dominated atmosphere, for a nitrogen-dominated atmosphere with the amount of hydrogen equal to the modern terrestrial level, and the hydrogen dominated atmosphere.}
  \label{f_avsp_3XUV}
\end{figure*}
\begin{figure*}
\includegraphics[width=2.0\columnwidth]{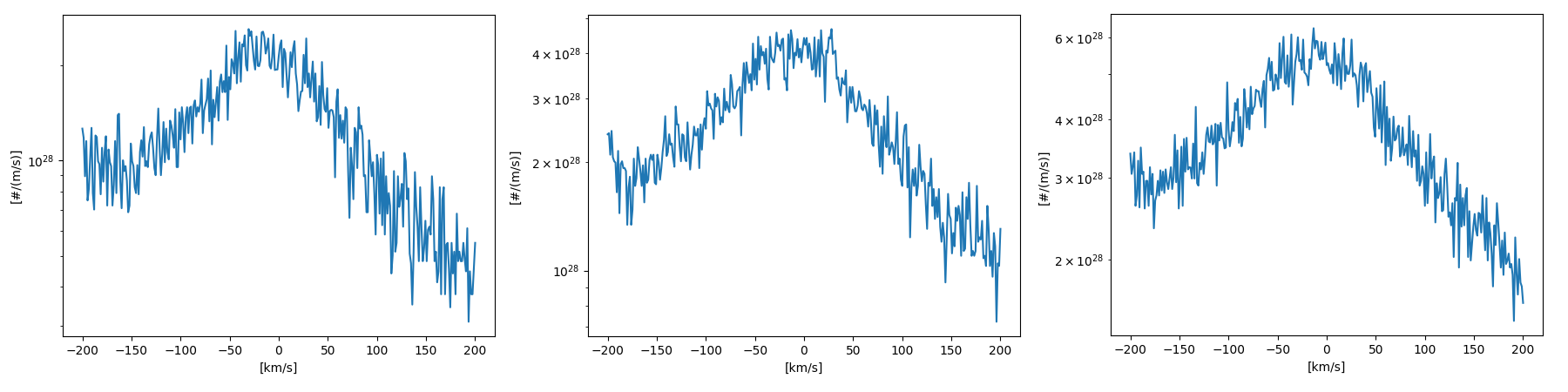}
 \caption{Velocity spectrum of neutral H atoms for a nitrogen-dominated atmosphere for 7, 10, and 20~XUV. Neutral hydrogen is produced only due to charge exchange. Originally, the newly produced hydrogen atoms have velocities of their stellar wind predecessors (a Maxwellian distribution with a maximum at 330~km/s), but then this distribution is altered by the collisions with other atmospheric neutrals. This is the source of the little ``bump'' in the velocity spectrum at very low velocities and the main source of neutral H particles in the part of the spectrum with the positive velocities. }
  \label{f_avsp_N}
\end{figure*}
\begin{figure*}
\includegraphics[width=2.0\columnwidth]{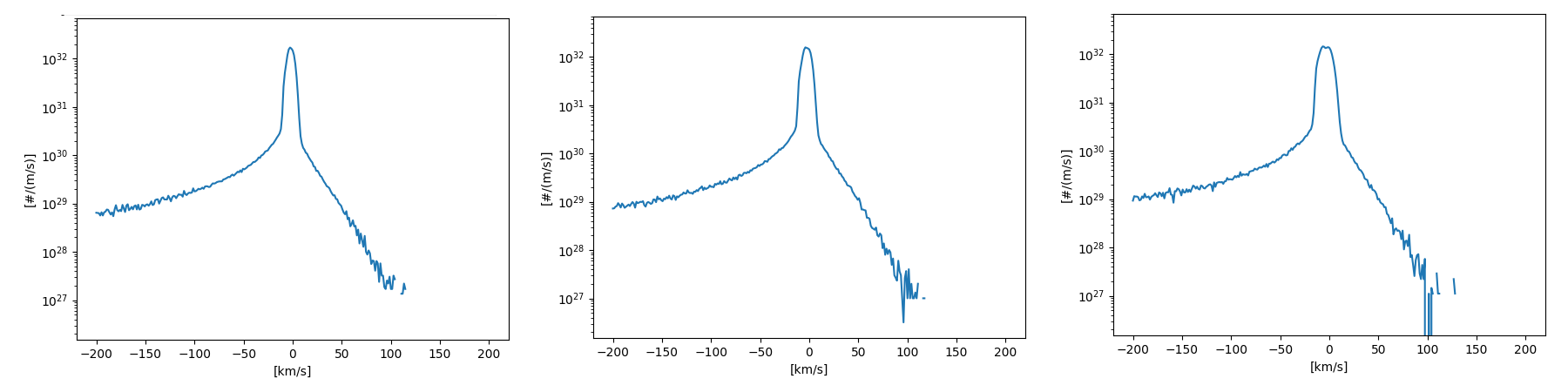}
 \caption{Velocity spectrum of neutral H atoms for a hydrogen-dominated atmosphere for 7, 10, and 20~XUV. Neutral hydrogen atoms at low velocities have predominantly  atmospheric origin, high-velocity particles at both negative and positive velocities are produced due to charge exchange with protons of the stellar wind. Similar to the hydrogen atoms distribution for nitrogen-dominated atmospheres, these velocity spectra are also altered by the collisions with the atmospheric neutrals with low velocities.}
  \label{f_avsp_H}
\end{figure*}

\begin{figure*}
\includegraphics[width=1.0\columnwidth]{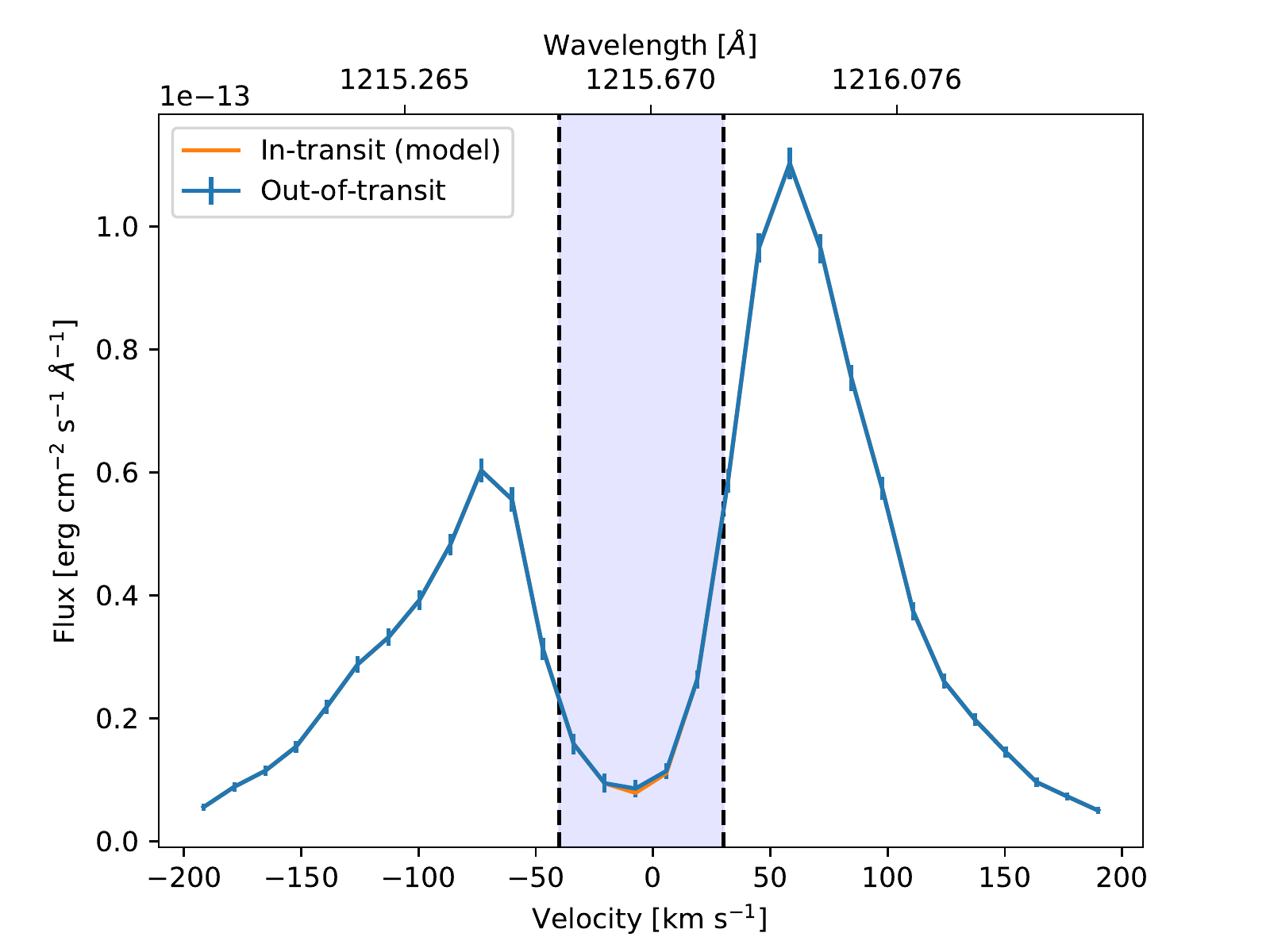}
\includegraphics[width=1.0\columnwidth]{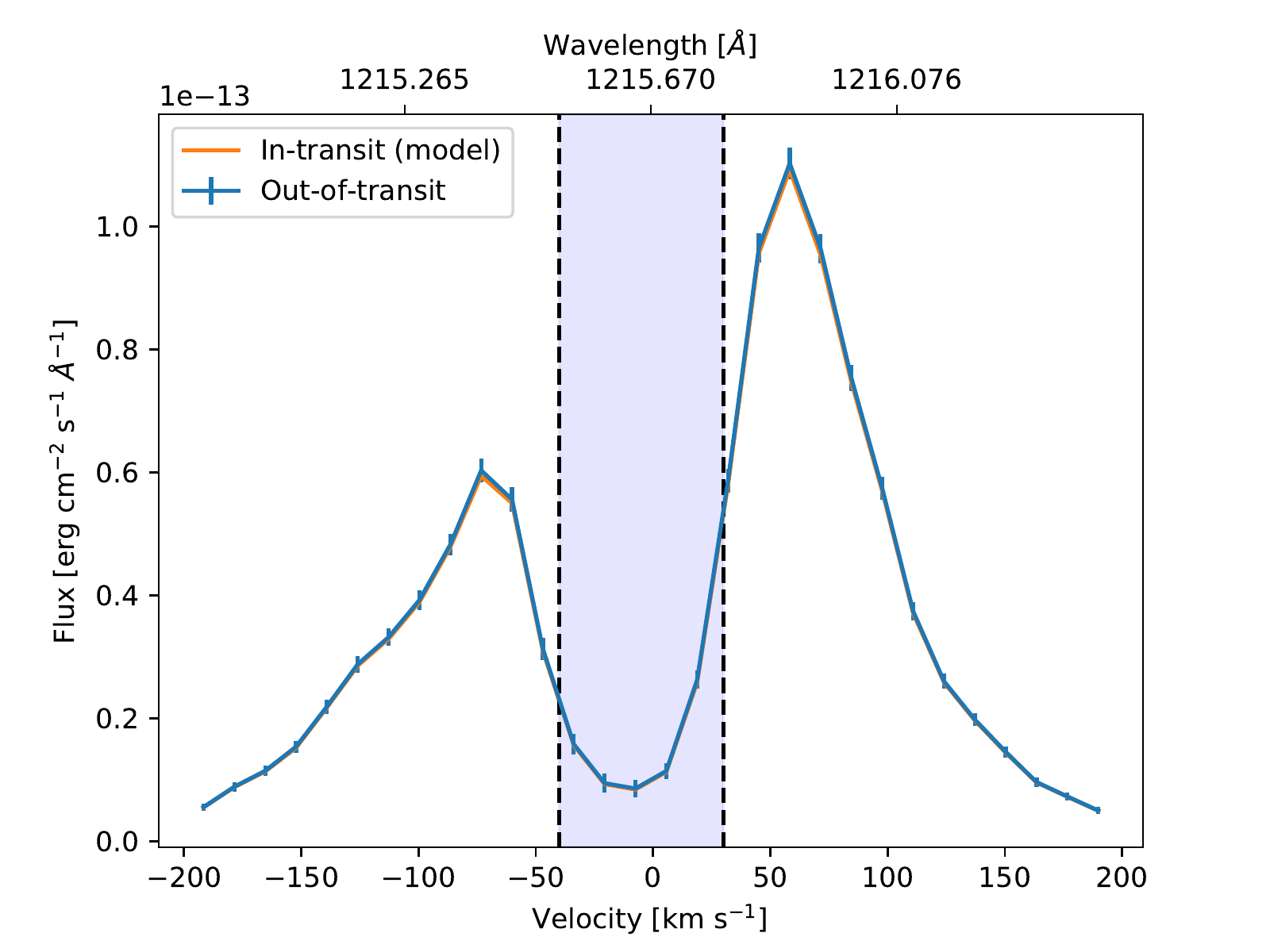}
 \caption{Modeled absorption for a nitrogen-dominated atmosphere at 3 XUV with a terrestrial hydrogen fraction (left) and a nitrogen-dominated atmosphere at 20 XUV (right). Modeled absorption is compared to the out-of-transit observation of GJ~436 \citep{Lavie17}. The blue line shows the observed out-of-transit Ly$\alpha$ profile of GJ~436. The orange line shows the modeled in-transit absorption. The region of the contamination by the geocoronal emission in the center of the line is marked by two vertical dashed lines. The observational errors are shown as errorbars of the blue line. One can see that in both cases the nitrogen-dominated atmospheres do not produce a significant in-transit Ly$\alpha$ signature, which is indicated by the fact that the orange line practically coincides with the blue line. }
  \label{f_AbsNit}
\end{figure*}

\begin{figure*}
\includegraphics[width=1.0\columnwidth]{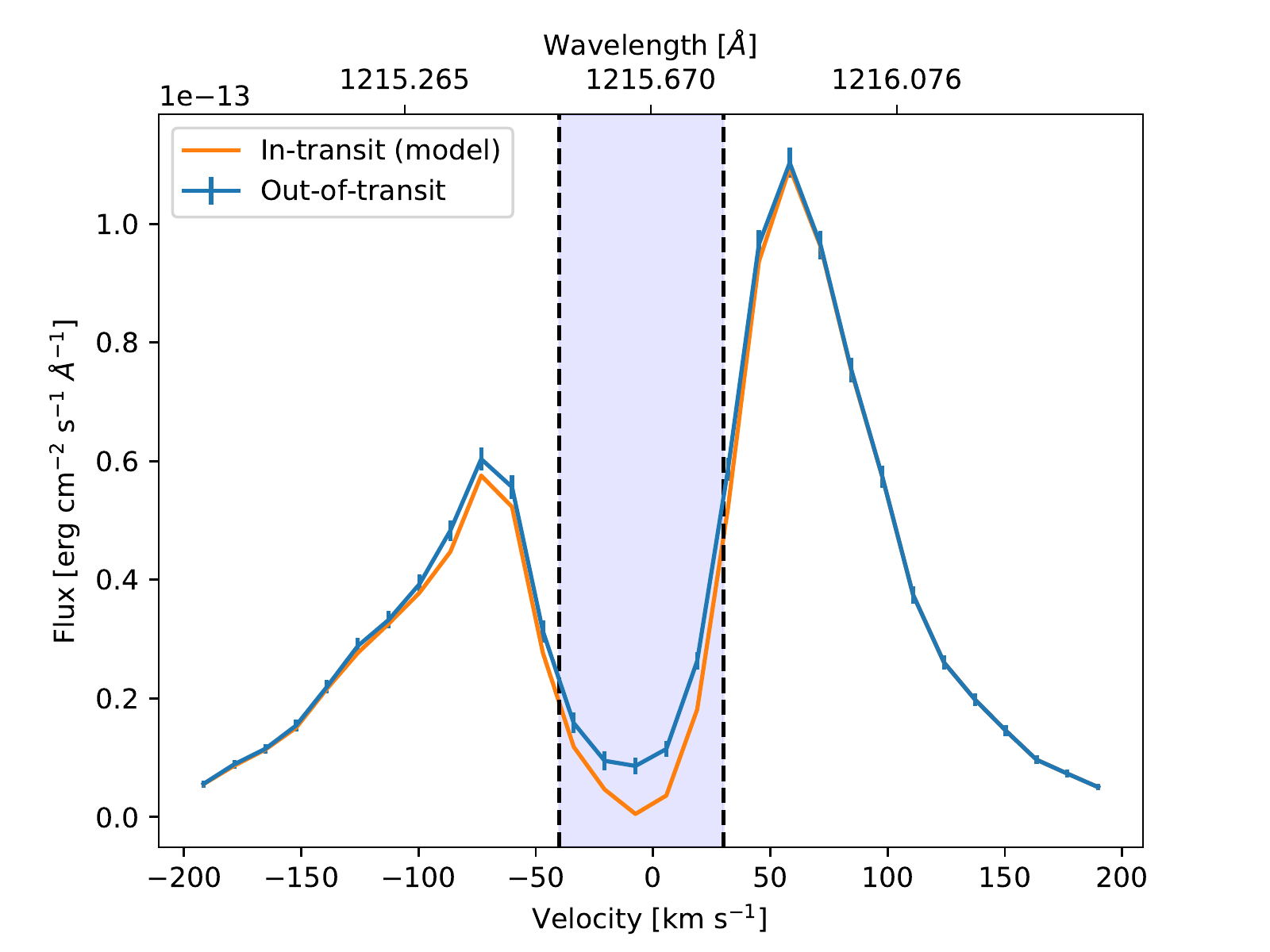}
\includegraphics[width=1.0\columnwidth]{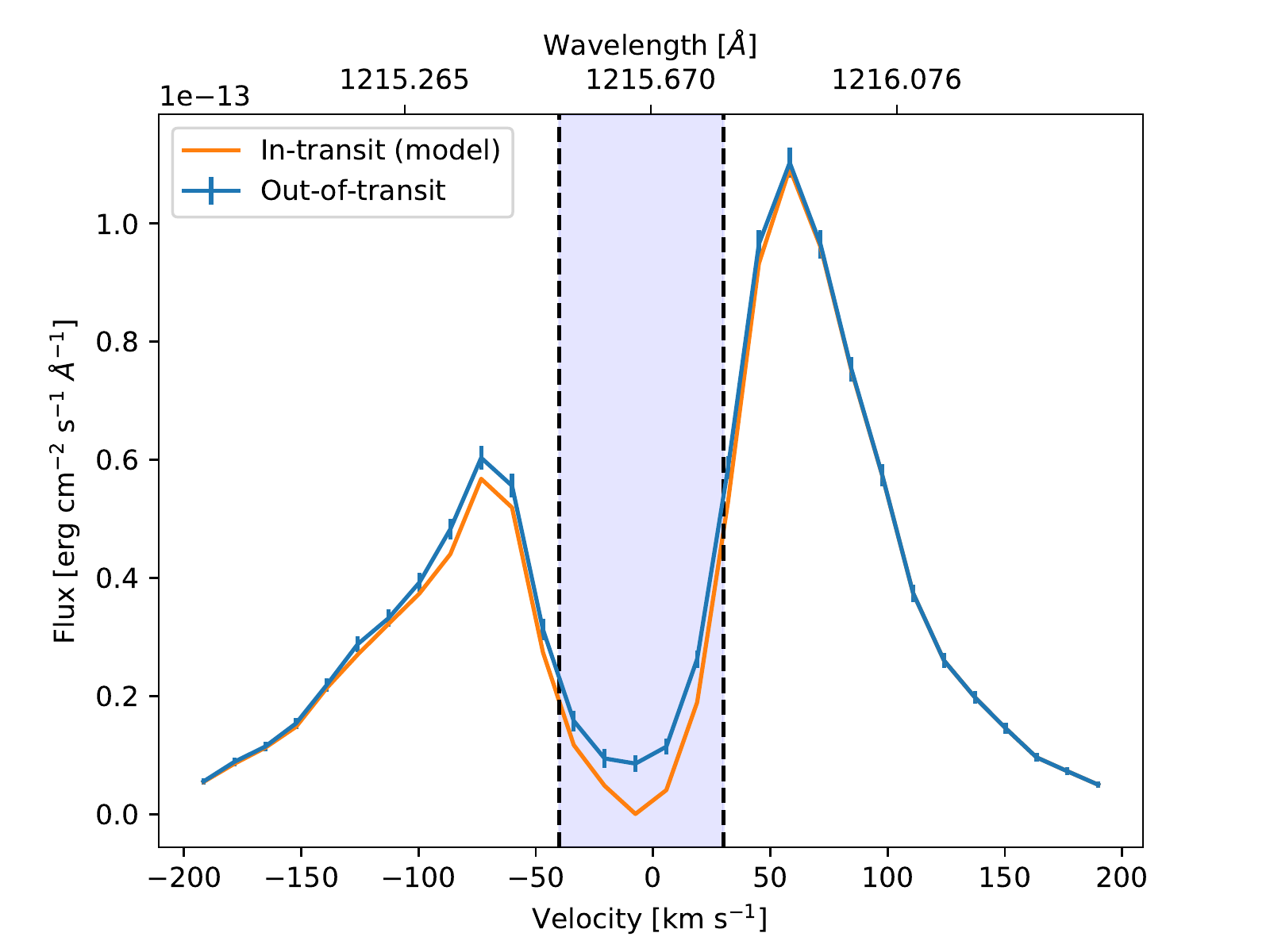}
\includegraphics[width=1.0\columnwidth]{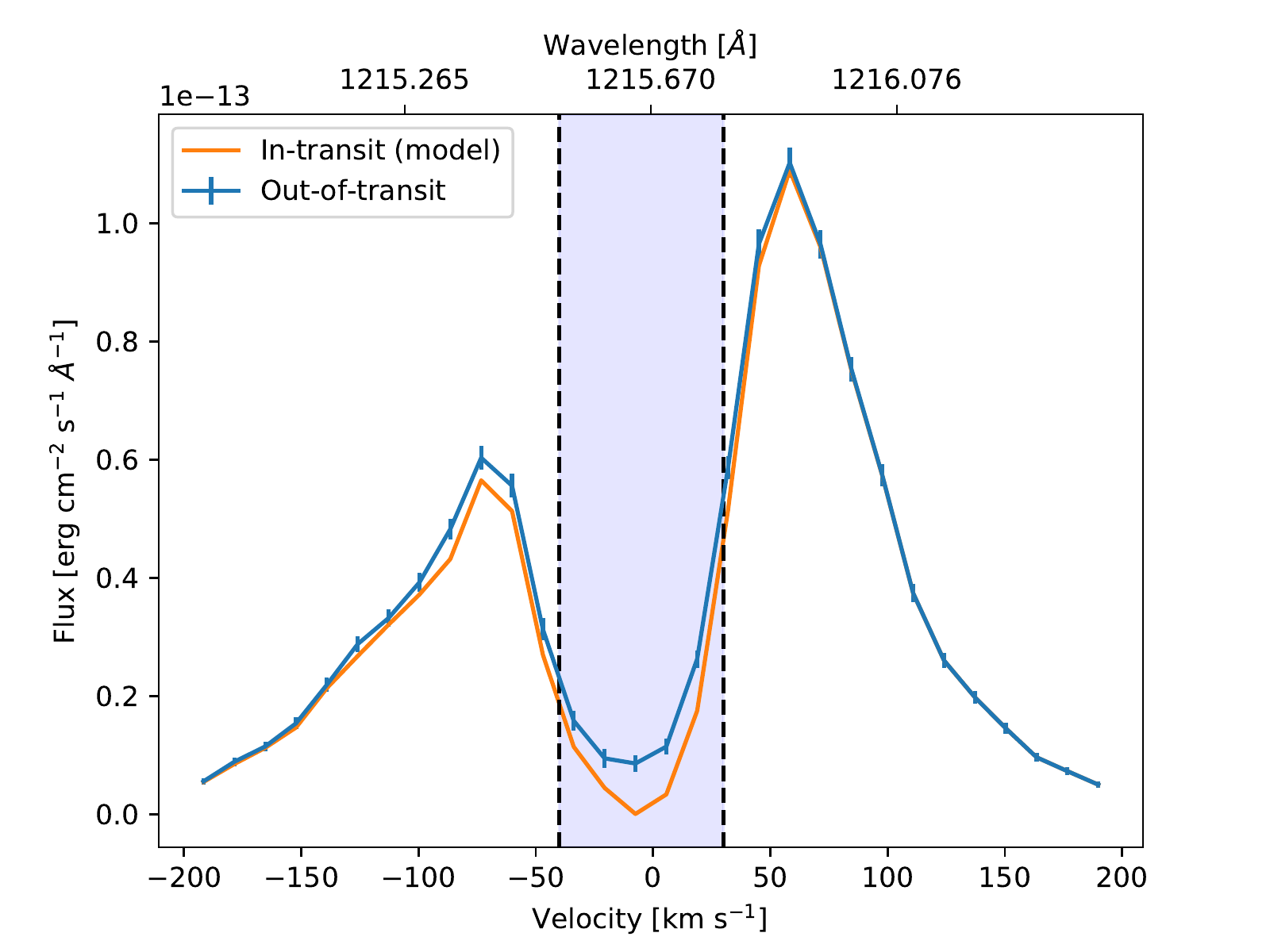}
\includegraphics[width=1.0\columnwidth]{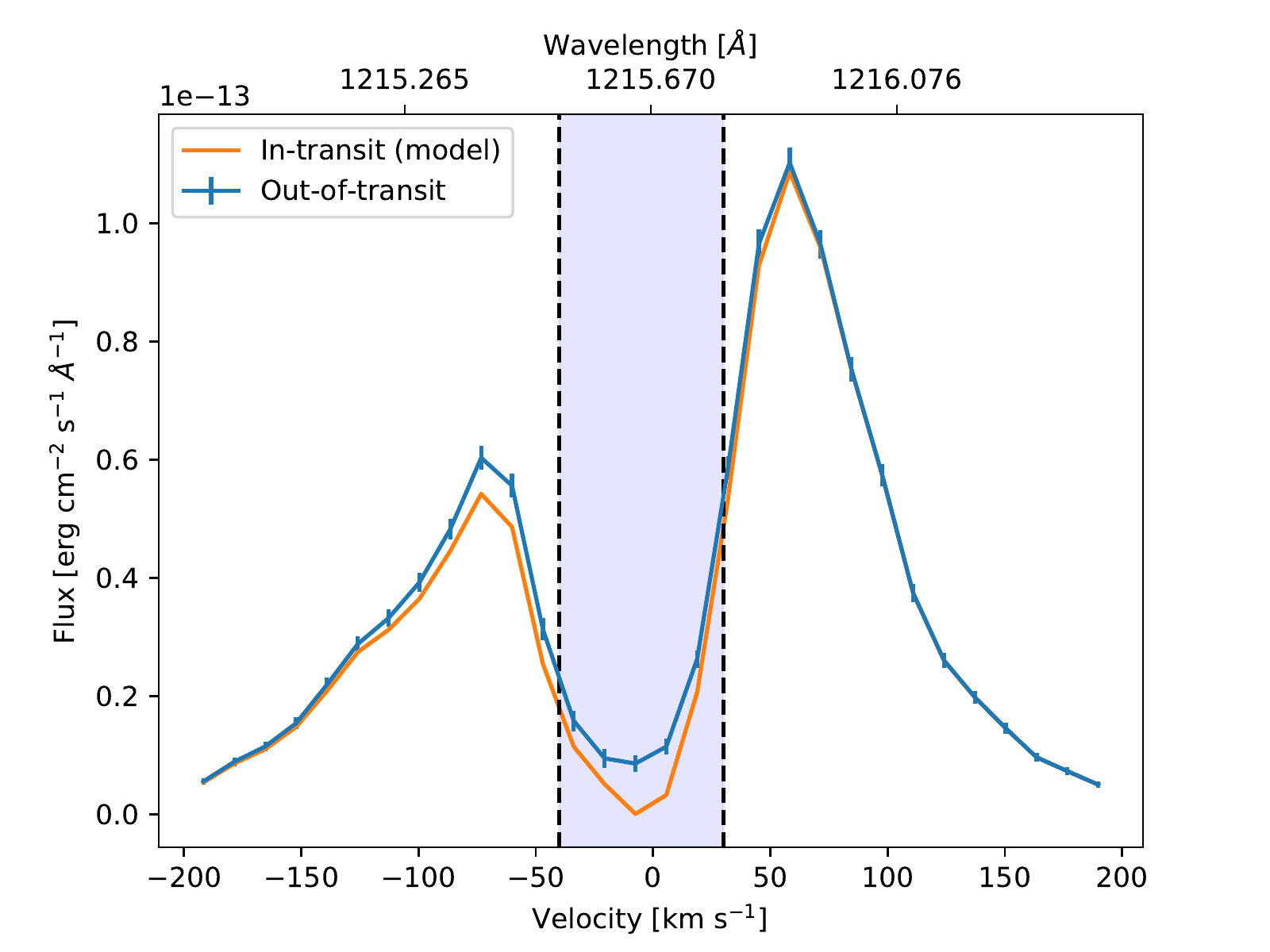}
 \caption{Same as Fig.~\ref{f_AbsNit}, but for a hydrogen-dominated atmosphere exposed to 3 (upper left), 7 (upper right), 10 (lower left) and 20 (lower right) times the current Earth XUV level. One can see that unlike a nitrogen-dominated atmosphere, hydrogen-dominated atmospheres produce a noticeable Ly$\alpha$ signature in all cases, however, the deepest in-transit absorption exceeding the observational errors in the blue wing of the line is obtained for the most expanded exospheres at 10 and 20~XUV. Although the red wing of the line also shows some absorption, it does not exceed the observational errors. }
   \label{f_AbsHyd}
\end{figure*}

Figs.~\ref{f_avsp_3XUV}, \ref{f_avsp_N}, and \ref{f_avsp_H} show velocity distributions for all H atoms in the simulations along the $x$-axis, which is directed toward the star. Fig.~\ref{f_avsp_3XUV} shows velocity spectra for the case of 3~XUV corresponding to clouds shown in Fig.~\ref{f_clouds3XUV}, from left to right: a nitrogen-dominated atmosphere, a nitrogen-dominated atmosphere with a terrestrial hydrogen fraction, and a hydrogen dominated atmosphere. Fig.~\ref{f_avsp_N} shows the H-atoms velocity distribution for a nitrogen-dominated atmosphere for 7, 10, and 20 XUV, and finally Fig.~\ref{f_avsp_H} shows the same for a hydrogen-dominated atmosphere. 

Comparing the left panel of Fig.~\ref{f_avsp_3XUV} to all cases shown in Fig.~\ref{f_avsp_N}, one can see that the amount of hydrogen particles produced due to charge exchange increases significantly between 3 and 7~XUV, which indicates a more intense interaction. This coincides with an increase in the cloud size, therefore, it seems reasonable to suggest that the increase in the amount of produced H~atoms is due to increase of the atmosphere cross section. Indeed, hydrodynamical models show that a nitrogen-dominated atmosphere with an Earth-like amount of CO$_2$ is very sensitive to the flux in the XUV radiation domain \citep{Tian08,Johnstone18}. If one compares all three panels of Fig.~\ref{f_avsp_N} to each other, one can see that the amount of neutral hydrogen is gradually increasing with increasing XUV fluxes.

Comparing Figs.~\ref{f_avsp_N} and \ref{f_avsp_H} to each other, one can see that the main central peak, which presents atmospheric H particles with small velocities and which is present for all hydrogen-dominated atmospheres is actually absent in case of nitrogen-dominated atmospheres, where one can see only the ENAs distribution. In Fig.~\ref{f_avsp_H} one can see that the central peak is actually slightly shifted toward the negative velocities. This is due to the combined effect of two factors, first, the radiation pressure, which accelerates hydrogen atoms away from the star, and second, due to enhanced ionization in the part of the cloud directed toward the star, where the particles are more likely to have positive velocities along the $x$-axis. In our model, ionization predominantly removes atoms with positive velocities, because electron impact ionization can occur to particles only outside the ionosphere, and photoionization is able to ionize only the particles outside the planetary shadow. This leads to some imbalanced removal of particles flying toward the star.

In all velocity spectra, one can see that the initial velocity of the ENAs is altered by collisions with other atmospheric neutrals. Initial ENAs velocities correspond to the stellar wind velocities, which present a Maxwellian distribution with a temperature of $2 \times 10^6$~K and a maximum at 330~km/s. On the other hand, Figs.~\ref{f_avsp_3XUV}--\ref{f_avsp_H} show that there is some population of neutral H atoms with the positive velocities (flying toward the star), as well as a small maximum (in case of nitrogen-dominated atmospheres) at small velocities. This is due to the fact that collisions lead to \textit{i)} isotropic scattering in a random direction and \textit{ii)} to some energy exchange between ENAs and atmospheric particles. 

 Figures~\ref{f_AbsNit} and \ref{f_AbsHyd} show our final results, namely, Ly$\alpha$ in-transit spectra compared to out-of-transit observation of GJ~436 \citep{Lavie17}. Simulated absorption along the line-of-sight shows similar features with the velocity spectra, as it is created by atoms with corresponding velocities. Fig.~\ref{f_AbsNit} shows in-transit absorption for a nitrogen-dominated atmosphere with the terrestrial hydrogen content at 3~XUV (left) and a nitrogen-dominated atmosphere at 20~XUV (right). As one can see, both cases reveal no significant in-transit absorption, even the case with a small amount of hydrogen present. One can see that it can produce a very weak absorption signature at low velocities, however, due to the low orbit of the Hubble Space Telescope this part of the spectrum is obscured by the geocorona and cannot be observed. The nitrogen-dominated atmosphere at 20~XUV produces some in-transit absorption due to ENAs in the blue and red wings of the Ly$\alpha$ line, however, the signal is almost negligible in comparison to the observational errors (shown as errorbars on the picture). We do not show simulated absorption for nitrogen-dominated atmospheres at 3, 7, and 10~XUV, because they produce even smaller signatures than the nitrogen-dominated atmosphere at 20~XUV, which are barely recognizable by eye. 
 
Finally, Fig~\ref{f_AbsHyd} summarizes our results for the in-transit absorption of hydrogen-dominated atmospheres. It shows calculated in-transit absorption signatures if a planet with a hydrogen-dominated atmosphere is exposed to 3 (upper left), 7 (upper right), 10 (lower left), and 20 (lower right) times the XUV flux of the modern Earth. As anticipated, absorption increases from 3 to 20 XUV. In the center of the Ly$\alpha$ line all atmospheres produce a deep absorption signature, however, this part of the line is contaminated by the geocorona. In the high velocity wings of the line, absorption increases with increasing XUV flux, being the strongest in the blue wing of the line at 20~XUV. One can see that in all cases the absorption exceeds the observational uncertainty and, therefore, can be considered detectable by the Hubble Space Telescope. However, the data by \citet{Lavie17} have very low observational errors because they combine several observational visits. If one compares the modeled absorption to errorbars for just one visit (for instance, one of the visits by \citealp{Ehrenreich15}), the in-transit signature only exceeds the observational errors for a hydrogen-dominated atmosphere exposed to 20~XUV.

\subsection{Influence of stellar wind parameters}
\label{sec_wind}

In this section, we investigated the influence of the assumed stellar wind parameters on our simulation results. The wind we assumed for the simulations corresponds to a mass-loss rate of $3.5 \times 10^{-14} M_{\odot}$~year$^{-1}$, which is approximately 2.5 times higher than the mass loss rate of the Sun, $\dot{M}_{\odot}$. This mass loss rate is approximately 30 times higher than the average value estimated for GJ~436 by \citet{Vidotto17}. This discrepancy can indicate that additional studies are necessary on the stellar wind models, ideally coupled with the Ly$\alpha$ observations of astrospheres and planets (e.g, \citealp{Wood05}). In this section, we performed two additional simulations to estimate the influence of the stellar wind on our results: first, we adopted the wind parameters from \citet{Vidotto17} and second, we tested the influence of a wind with lower temperature, lower velocity, and larger density than the ones assumed above. We assumed a hydrogen atmosphere exposed to 20 times the modern Earth's XUV flux, with the deepest simulated in-transit Ly$\alpha$ absorption, as a proxy. 

\noindent\textbf{Stellar wind according to \citet{Vidotto17}.} For this simulation, we adopted the stellar wind proton density and velocity for the habitable zone from Fig.~3 of \citet{Vidotto17}: 

\begin{itemize}
	\item Stellar wind density $n_{sw} = 1.0 \times 10^7$~m$^{-3}$;
	\item Stellar wind velocity $v_{sw} = 270$~km/s;
	\item Stellar wind temperature $T_{sw} = 3.95 \times 10^5$~K. 
\end{itemize}

\begin{figure*}
\includegraphics[width=1.0\columnwidth]{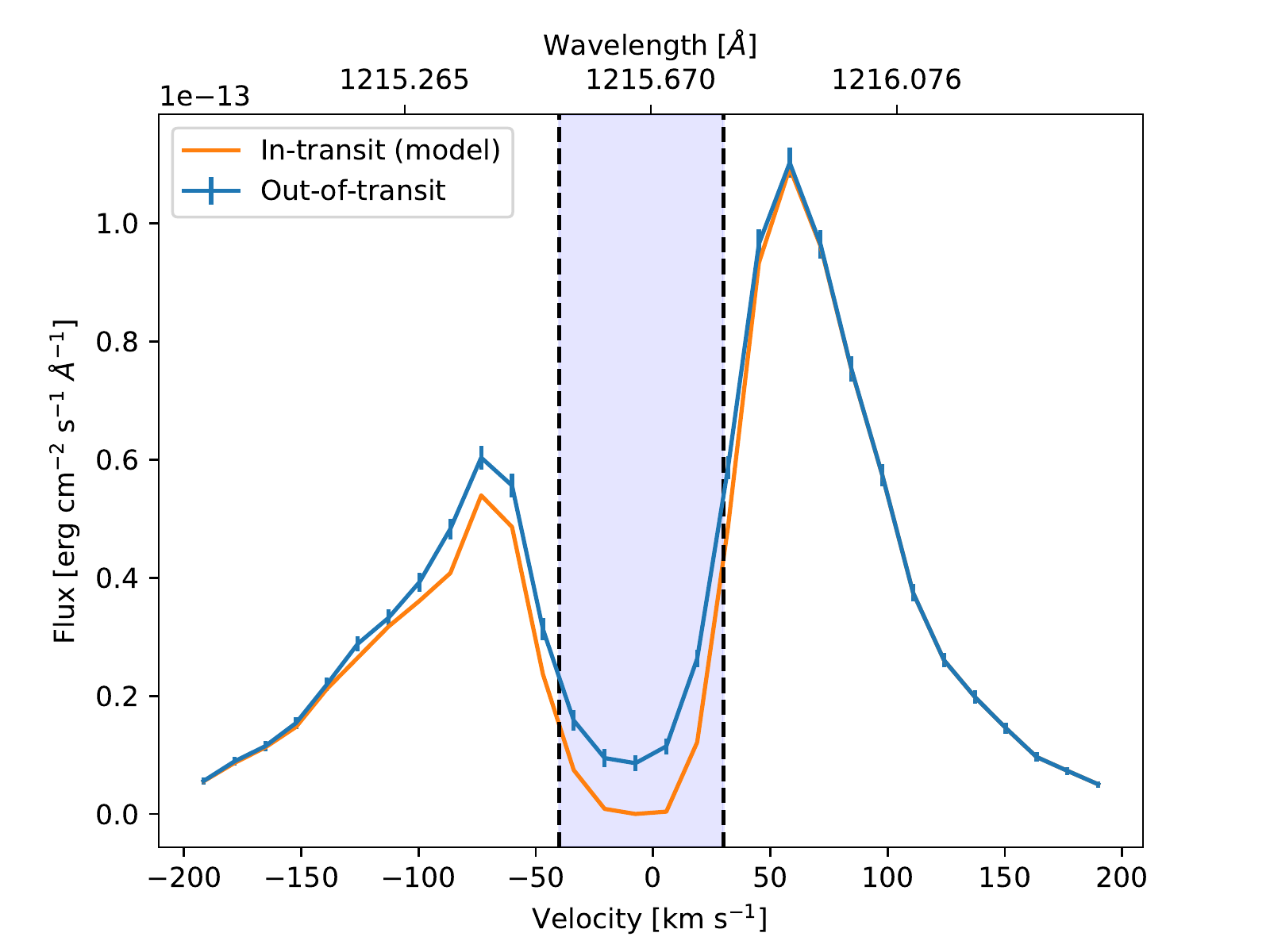}
\includegraphics[width=1.0\columnwidth]{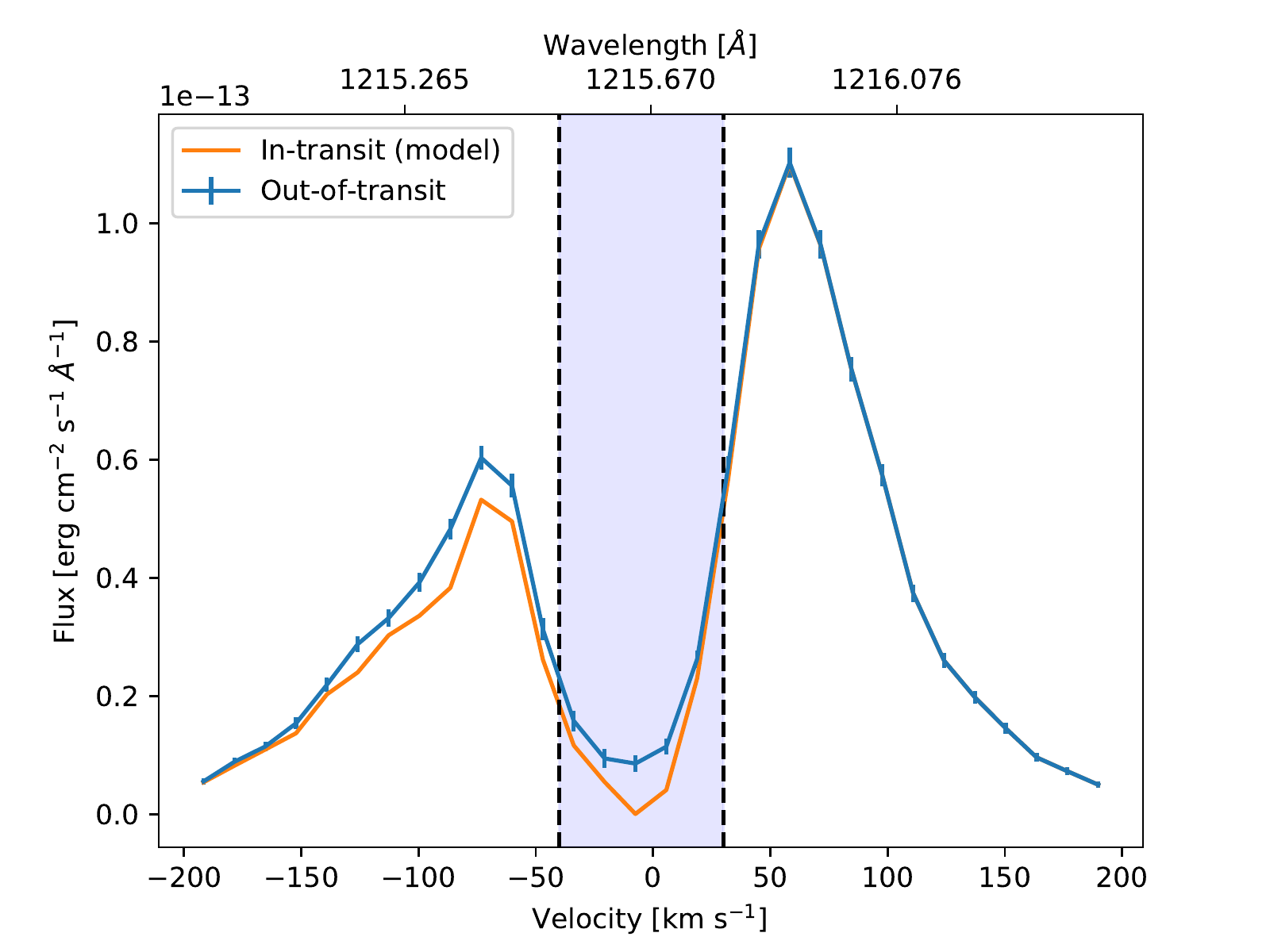}
 \caption{Same as Fig.~\ref{f_AbsHyd} (lower right panel, hydrogen atmosphere exposed to 20 times Earth's XUV level), but for different stellar wind parameters. Left panel: stellar wind parameters according to \citet{Vidotto17}. Right panel: stellar wind with lower temperature, lower velocity, and larger density. See text for details.}
   \label{f_AbsAltSW}
\end{figure*}

The temperature is adopted from \citet{K13}, as before. For this simulation, we changed the stellar wind electron impact ionization rate according to the lower wind density to $\tau_{ei} = 2.7 \times 10^{-7}$~s$^{-1}$. This stellar wind corresponds to a mass-loss rate of $1.2 \times 10^{-15} M_{\odot}$~year$^{-1}$. The left panel of Fig.~\ref{f_AbsAltSW} shows modeled in-transit absorption in the Ly$\alpha$ line. The absorption due to the ENAs decreased due to a lower flux of the stellar wind. The absorption due to atmosphere broadening almost did not change in comparison to the default 20~XUV case (Fig.~\ref{f_AbsHyd}, lower right panel), although it is marginally larger due to lower electron impact ionization.

\noindent\textbf{Simulation for a denser and slower stellar wind.} For this case, we assumed the following parameters:

\begin{itemize}
	\item Stellar wind density $n_{sw} = 8.25 \times 10^8$~m$^{-3}$;
	\item Stellar wind velocity $v_{sw} = 100$~km/s;
	\item Stellar wind temperature $T_{sw} = 4.0 \times 10^5$~K. 
\end{itemize}

These parameters still correspond to the same mass loss rate as in the standard stellar wind used in the simulations. Stellar wind electron impact ionization rate was assumed equal to $\tau_{ei} = 2.3 \times 10^{-5}$~s$^{-1}$. Since a denser wind mowing with lower velocity can produce more ENAs in the observed velocity domain (-200 $\le v_x \le$ 200~km/s), one could expect a deeper absorption signature in comparison to the default case. However, right panel of Fig.~\ref{f_AbsAltSW} shows that this is not the case. This counterintuitive result is explained by the fact that the electron impact ionization rate is proportional to the wind density and is, therefore, higher for a denser wind. For this reason, although more ENAs are produced due to charge exchange with a denser wind, also more hydrogen atoms are ionized, which counterbalances the ENAs effect on the Ly$\alpha$ transit signature, and the absorption is only marginally increased.

Fig.~\ref{f_AbsAltSW} shows qualitatively the same absorption as Fig.~\ref{f_AbsHyd} for the 20~XUV case. This indicates that in the considered case the Ly$\alpha$ absorption is to a large extent influenced by the planetary atmosphere, and to a lesser extent by the ENAs and the interaction with the stellar wind. A denser stellar wind can also influence the absorption of magnetized planets, because through its RAM pressure it is influencing the location of the substellar point and the general magnetosphere shape which, in turn, determines which portion of the atmospheric neutrals is exposed for the interaction with the wind (see, e.g., \citealp{K14b} for discussion). However, the study of magnetized planets is beyond the scope of the present article.

\section{Discussion}
\label{sec_disc}

As we have shown above, at the distance of $\sim$10~pc (the distance to GJ~436) only hydrogen-dominated atmospheres would be detectable around Earth-like planets in the HZ of M~dwarfs using HST, and only in case of high XUV fluxes or increased observation times for stars with lower activities. Nitrogen-dominated atmospheres are non-detectable even at 20~XUV. According to our results, an Earth-like exosphere content also does not produce an observable Ly$\alpha$ signature. This result would likely be the same for other secondary atmospheres such as dominated by carbon dioxide. In fact, since CO$_2$ is an efficient thermosphere coolant, it effectively decreases the temperature of the upper atmosphere thus protecting it from the expansion \citep{Johnstone18}. A less expanded atmosphere would present a smaller cross section for the interaction with the stellar wind, therefore reducing the amount of hydrogen atoms produced by charge exchange reactions. Taking this into account, we can conclude that at 10 pc and beyond HST is able to detect only hydrogen-dominated atmospheres. This is in agreement with recent observations including non-detections of Ly$\alpha$ signatures by HST on small, presumably rocky, exoplanets \citep{Ehrenreich12,B17}, possibly indicating that these planets either possess secondary (dominated by other species than hydrogen) atmospheres or are airless bodies due to efficient non-thermal escape. As we have shown in this paper, HST is not able to distinguish between these two possibilities for a target at 10~pc. A more advanced instrument such as LUVOIR or WSO UV (due to its higher orbit above the geocorona, which would allow to detect also the core of the Ly$\alpha$ line) might be able to do it. 

One should keep in mind that we always compare the modeled absorption to the observed out-of-transit Ly$\alpha$ line of GJ~436, which corresponds to the XUV flux in the HZ of $\approx$3~XUV. On the other hand, our simulations have been performed for various XUV enhancement factors (3, 7, 10, and 20~XUV). If the star would be more
active in XUV its Ly$\alpha$ line would also be stronger (see Fig.~\ref{f_LyaGJ436}), in which case the errorbars of the observations would likely be smaller for the
same observational setup.

Recently, \citet{Castro18} also studied if exospheres of Earth-like planets in the HZs of M~dwarfs can be observable by World Space Observatory UV (WSO UV). They did not perform a modeling of the upper atmosphere and used a more simplified approach, however, still arriving at similar conclusions. For an Earth-like exosphere (a nitrogen-dominated upper atmosphere with a small hydrogen content) and an Earth-like XUV flux they have assumed a modified barometric formula which accounts for a zenith angle. For higher XUV fluxes, they have adopted a hydrogen-dominated 1D atmosphere profile from \citet{Erkaev13}, calculated using a similar version of the code we used to obtain atmospheric profiles for hydrogen-dominated atmospheres in the current article. For both cases, \citet{Castro18} did not take into account the interaction between the stellar wind and the neutral atmosphere. Even if a planet possesses a magnetic field, it may not protect the atmosphere from the dense and fast stellar wind of M~dwarfs, which can sufficiently compress the atmosphere (e.g., \citealp{Vidotto13}) and lead to an interaction pattern similar to the one of a non-magnetized planet considered in this article.

Despite different approaches, our results and the results of \citet{Castro18} are in agreement. According to \citet{Castro18}, an Earth-like exosphere can be detected only as close as 1.35~pc, which is only slightly farther than the distance to the nearest star to the Solar system, Proxima~Centauri (1.295~pc). This agrees with our conclusion that Ly$\alpha$ signatures of nitrogen-dominated atmospheres of a planet orbiting in the HZ of GJ~436 at 10.2~pc would be non-detectable by the HST. On the other hand, \citet{Castro18} concluded that hydrogen-dominated atmospheres may be detectable by LUVOIR at further distances of about 10~pc, similar to our conclusion that HST may detect a hydrogen-dominated atmospheres of an Earth-like planet in the HZ of a star similar to GJ~436.

Therefore, there is a possibility that HST might be able to observe Earth-like exospheres at very close distances, such as that of Proxima~Centauri. However, Proxima~Centauri~b is not a transiting planet \citep{Anglada16}, thus rendering these observations impossible. It is possible that the HST can detect hydrogen-dominated atmospheres around other nearby M~dwarfs  with low XUV fluxes within 5--10~pc. The TESS satellite, which is an all-sky survey which will detect transiting exoplanets orbiting the closest M~dwarfs, may provide promising targets for observations by HST and, in the future, LUVOIR and WSO~UV.

A recent study \citep{dosSantos18} estimated the observability of an Earth-like exosphere by LUVIOUR, coming to the conclusion that an Earth-like geocorona would be detectable by LUVOIR for planets orbiting stars not further than about 15~pc. They also arrived at the conclusion that at such distances the geocorona would not be detectable by the HST, which is in full agreement with our conclusions. In the current study, we have additionally considered different atmospheric structures and XUV enhancement factors, showing that the HST may detect a hydrogen atmosphere of an Earth-like planet orbiting an active nearby star, thus indicating that such observations are possible at present with the HST. However, our conclusions also point out that only future high-precision instruments such as LUVOIR could detect a hydrogen corona similar to Earth's geocorona.


In this paper, we assumed quite large elastic collision cross sections and an isotropic scattering of atoms after a collision. In reality, cross sections decrease with increasing energy of the particles (e.g., \citealp{Izmodenov00}) and the direction of the scattering is also non-isotropic \citep{Izmodenov00,Lindsay05b}. This simplified assumption leads to an overestimate of ENAs in the red part of the Ly$\alpha$ line, because these particles with positive velocities (flying toward the star) are produced due to scattering on the neutral atmospheric particles. However, as we have shown in our results, for a terrestrial planet even in this case the ENAs do not produce a significant Ly$\alpha$ signature in the red part of the spectrum. Therefore, different collision cross sections would not change our conclusions. Implementing a velocity-dependent collision cross section and also taking into account the dependence of the scattering direction on the impact angle will be taken into account in future studies. 

In our model, we did not take into account the influence of the interplanetary magnetic field (IMF) on the planetary ionosphere, therefore, our results are valid for a case of a small IMF, when the protons of the stellar wind can freely penetrate the planetary ionosphere. IMF can form a pile up region near the ionosphere \citep{Erkaev17}, which will prevent the stellar wind from the interaction with the ionosphere, thus making the interaction region and the amount of generated ENAs smaller. However, since we were interested in estimating the Ly$\alpha$ signatures for the maximum interaction, but still did not find any significant absorption in the Ly$\alpha$ line for the nitrogen atmospheres, our main conclusions would not change for a case with a non-negligible IMF. For hydrogen-dominated atmospheres, the presence of an IMF may reduce the number of ENAs and the absorption in the left part of the spectrum, however, it would not affect the absorption caused by broadening of the spectral line by the atmosphere, or possibly even increase it due to decreased ionization by charge exchange and electron impacts.

We also do not take into account the effects of the formation of a bow shock and assume that stellar wind density and velocity do not change near the planetary ionospheric obstacle. Bow shock formation can lead to deceleration and temperature and density increase of the stellar wind near the planet. ENAs are mostly formed in the region between the ionopause and bowshock \citep{Khodachenko17}. Taking into account these effects would slightly increase the amount of ENAs inside the observable velocity domain (from -200 to 200~km/s), but it is unlikely that even an increased ENA contribution would change our conclusion about observability of the nitrogen atmospheres in the Ly$\alpha$ line. As for hydrogen atmospheres, although ENAs can be a substantial contributor to the absorption signature, for the considered cases most of the absorption is produced due to line broadening, similar to hot Jupiters \citep{Shaikhislamov16}. Therefore, it is unlikely that bow shock effects would alter our conclusions.

As a final note, we would like to repeat that nitrogen-dominated atmospheres may not be very common on planets orbiting M~dwarfs. As we have stated before, nitrogen-dominated atmospheres with low CO$_2$ ratios are subject to enhanced expansion and non-thermal losses if exposed to XUV levels higher than 5--7 \citep{Tian08,Lichtenegger10}. Even though GJ~436 does not exhibit a high level of XUV radiation in its habitable zone at present, it may have been much more active in the past. An increased level of CO$_2$ may prevent the atmosphere from expansion and thus protect it \citep{Kulikov07,Lichtenegger10,Johnstone18}, but this would require much higher CO$_2$ levels in comparison to the present-day Earth, or a very inactive M~dwarf thorough its history. Another argument which would suggest that nitrogen-dominated atmospheres may be rather rare is that Earth's, its nitrogen atmosphere is sustained also due to plate tectonics and, indirectly, via influence of biologic activity \citep{L18}.

\section{Conclusions}
\label{sec_conc}

In this paper, we have performed modeling of the exospheres of an exoplanet orbiting in the HZ of GJ~436. We have assumed that the planet has a radius and a mass equal to those of the the Earth, and have considered hydrogen and nitrogen dominated atmospheres exposed to different XUV fluxes. We have applied sophisticated modeling which included 1D lower atmosphere models and a multi-species DSMC upper atmosphere code, and modeled atomic coronae surrounding the planets at 3, 7, 10, and 20~XUV. At 3~XUV, which is the current level of short-wavelength radiation in the middle of the HZ of GJ~436, we also additionally modeled a nitrogen-dominated atmosphere with an Earth-like hydrogen content. For all modeled atomic coronae, we have calculated the absorption in the Ly$\alpha$ line that a transit of such an atmosphere would produce, and compared it to the observed out-of-transit spectrum of the Ly$\alpha$ line of GJ~436 \citep{Lavie17}. In case of hydrogen-dominated atmospheres, the signature in the Ly$\alpha$ line is produced due to the combined effects of spectral line broadening in the massive planetary atmosphere and the energetic neutral atoms generated due to charge exchange between stellar wind and the planetary neutrals. In case of a secondary atmosphere, which we assumed to be dominated by atomic nitrogen, the Ly$\alpha$ signature is produced due to ENAs alone. 

According to our results, only hydrogen-dominated atmospheres of an Earth-like planet can be detected by the HST around GJ~436. At 3~XUV, a hydrogen-dominated atmosphere can be detected by the HST in case of multiple observations. Closer or more active M~dwarfs (otherwise similar) will have stronger Ly$\alpha$ lines, therefore the SNR of the observations may also be better and merging multiple visits may not be necessary. Therefore, we conclude that hydrogen atmospheres may be detectable if the atmosphere of a planet is extended thus producing a very spacious exosphere. This is the case for close-in planets with sufficient hydrogen amount, such as hot Neptunes and Jupiters. We have reproduced the Ly$\alpha$ observations of GJ~436b assuming an escaping hydrogen-dominated atmosphere with a size significantly larger than the one of a terrestrial planet. Unlike terrestrial planets modeled in this article, GJ~436b produces a deep absorption signature in the Ly$\alpha$ line. 

Our results indicate that further observations in the Ly$\alpha$ line can be more fruitful for expanded hydrogen-dominated planets orbiting active M~dwarfs such as GJ~436b. Planets possessing secondary atmospheres do not produce a signature in the Ly$\alpha$ line strong enough to be detected by the Hubble Space Telescope at 10~pc or beyond, however, Ly$\alpha$ signatures of Earth-like planets with hydrogen-dominated atmospheres can be detected. It is possible that Ly$\alpha$ signatures of secondary atmospheres may be observable in the future using more advanced instrumentation, such as LUVOIR and WSO~UV.

\begin{acknowledgements}

We acknowledge the support by the Austria Science Fund (FWF) NFN project S116-N16 and the subprojects S11607-N16, S11606-N16 and S11604-N16. PO, HL, and NVE acknowledge support from the Austrian Science Fund (FWF) project P25256-N27 ``Characterizing Stellar and Exoplanetary Environments via Modeling of Lyman-$\alpha$ Transit Observations of Hot Jupiters''. NVE also acknowledges support by the RFBR grant No 16-52-14006. MLK also acknowledges FWF projects I2939-N27 and the partial support by the Ministry of Education and Science of Russian federation (Grant No. RFMEFI61617X0084). IFS acknowleges support of Russian Science Foundation project 18-12-00080.
The software used in this work was in part developed by the DOE NNSA-ASC OASCR Flash Center at the University of Chicago. This research was conducted using resources provided by the Swedish National Infrastructure for Computing (SNIC) at the High Performance Computing Center North (HPC2N), Ume$\mathring{\rm a}$ University, Sweden. 

The authors are very thankful to Dr. David Ehrenreich for providing the Ly$\alpha$ spectra of GJ~436b, which were used in this article. We would also like to sincerely thank Dr. Vincent Bourrier and Baptiste Lavie for original processing of these spectra.

\end{acknowledgements}

\bibliographystyle{aa} 

\end{document}